\documentclass{amsproc}
\usepackage{amsmath, amsthm, amssymb, slashed, framed}
\usepackage{ifpdf}
\ifpdf
  \usepackage[pdftex]{graphicx}
  \usepackage{epstopdf}
\else
  \usepackage[dvips]{graphicx}
\fi
\usepackage{url}
\newtheorem{theorem}{Theorem}[section]

\theoremstyle{definition}

\newtheorem{example}[theorem]{Example}

\newtheorem{conjecture}[theorem]{Conjecture}

\theoremstyle{remark}

\numberwithin{equation}{section}

\title{Quantization via Mirror Symmetry}

\author{Sergei Gukov}
\address{California Institute of Technology, Pasadena, CA 91125, USA\\
\newline
Max-Planck-Institut f\"ur Mathematik, Vivatsgasse 7, D-53111 Bonn, Germany.}
\email{gukov@theory.caltech.edu}
\thanks{
Prepared for the Takagi Lectures 2010.}

\font\teneurm=eurm10 \font\seveneurm=eurm7 \font\fiveeurm=eurm5
\newfam\eurmfam
\textfont\eurmfam=\teneurm \scriptfont\eurmfam=\seveneurm
\scriptscriptfont\eurmfam=\fiveeurm

 \font\teneusm=eusm10 \font\seveneusm=eusm7 \font\fiveeusm=eusm5
\newfam\eusmfam
\textfont\eusmfam=\teneusm \scriptfont\eusmfam=\seveneusm
\scriptscriptfont\eusmfam=\fiveeusm

\font\tencmmib=cmmib10 \skewchar\tencmmib='177
\font\sevencmmib=cmmib7 \skewchar\sevencmmib='177
\font\fivecmmib=cmmib5 \skewchar\fivecmmib='177
\newfam\cmmibfam
\textfont\cmmibfam=\tencmmib \scriptfont\cmmibfam=\sevencmmib
\scriptscriptfont\cmmibfam=\fivecmmib
\def\cmmib#1{{\fam\cmmibfam\relax#1}}

\def\example#1{\bgroup\narrower
\baselineskip\footskip\bigbreak
\hrule\medskip\nobreak\noindent {\bf Example}. {\it #1\/}\par\nobreak}
\def\endexample{\medskip\nobreak\hrule\bigbreak\egroup}

\newcommand{\be}{\begin{equation}}
\newcommand{\ee}{\end{equation}}

\newcommand{\C}{\mathbb{C}}
\newcommand{\Z}{\mathbb{Z}}
\newcommand{\R}{\mathbb{R}}

\newcommand{\IH}{\mathbb{H}}

\newcommand{\g}{{\mathfrak g}}
\def\neg{\negthinspace}
\newcommand{\LG}{{}^L\neg G}

\def\dual#1{{{}^L\negthinspace #1}}

\def\frak{\mathfrak}

\def\tilde{\widetilde}

\def\bar{\overline}

\def\CA{{\mathcal A}}
\def\CB{{\mathcal B}}

\def\CD{{\mathcal D}}
\def\CE{{\mathcal E}}

\def\CH{{\mathcal H}}
\def\CI{{\mathcal I}}

\def\CL{{\mathcal L}}
\def\CM{{\mathcal M}}
\def\CN{{\mathcal N}}
\def\CO{{\mathcal O}}
\def\CP{{\mathcal P}}

\def\CT{{\mathcal T}}

\def\CW{{\mathcal W}}

\def\CZ{{\mathcal Z}}

\def\EUM{\mathcal M}
\def\M{{\EUM}}
\def\MH{{{\EUM}_H}}

\def\BB{\cmmib B}  
\def\FF{\cmmib F}  

\def\Bcc{{\mathcal B}_{cc}}
\def\tCB{\widetilde{\mathcal B}}
\def\tBcc{{\widetilde{\mathcal B}}_{cc}}
\def\cp{{\mathbb{C}}{\mathbf{P}}}

\def\leadsto{\rightsquigarrow}
\def\p{\partial}

\DeclareMathOperator{\Ext}{Ext}

\DeclareMathOperator{\ch}{ch}
\DeclareMathOperator{\Tr}{Tr}
\DeclareMathOperator{\tr}{tr}

\DeclareMathOperator{\diag}{diag}
\DeclareMathOperator{\rank}{rank}

\begin{document}

\subjclass{}
\date{November, 2010}

\begin{abstract}
When combined with mirror symmetry, the $A$-model approach to quantization
leads to a fairly simple and tractable problem.
The most interesting part of the problem then becomes finding the mirror
of the coisotropic brane.
We illustrate how it can be addressed in a number of interesting examples
related to representation theory and gauge theory,
in which mirror geometry is naturally associated with the Langlands dual group.
Hyperholomorphic sheaves and $(B,B,B)$ branes play an important role
in the $B$-model approach to quantization.
\end{abstract}

\maketitle

\tableofcontents


\section{Introduction}

\hfill{\vbox{\hbox{\it Anyone who is not shocked by quantum theory}
\hbox{\it has not understood a single word.}}}

\hfill{\vbox{\hbox{Niels Bohr}}}

\medskip


The quantization problem of a symplectic manifold $(M,\omega)$
can be approached via the topological $A$-model of $Y$, a complexification of $M$ \cite{GW}.
In this approach, the Hilbert space $\CH$ obtained by quantization of $(M,\omega)$
is the space of open string states between two $A$-branes, $\CB'$ and $\Bcc$,
\be
\CH \; = \; {\rm space~of~} (\Bcc,\CB') {\rm ~strings} \,,
\label{hphys}
\ee
where $\CB'$ is an ordinary Lagrangian $A$-brane,
and $\Bcc$ is a space-filling coisotropic $A$-brane.
More formally, we can write \eqref{hphys} as the space of morphisms
\be
\CH \; = \; {\rm Hom} (\Bcc,\CB')
\label{hmath}
\ee
between two objects, $\Bcc$ and $\CB'$, in the Fukaya category of $Y$.

In general, in a Fukaya category 
the space of morphisms between two Lagrangian objects, $\CB_1$ and $\CB_2$,
is given by the symplectic Floer homology, $HF_{{\rm symp}}^* (\CB_1, \CB_2)$.
Therefore, if both of our objects in \eqref{hmath} were familiar
Lagrangian objects, the space of morphisms $\CH$ would be obtained
by counting their intersection points and analyzing pseudo-holomorphic
disks with boundary on $\CB_1$ and $\CB_2$.

However, our situation is more complicated and more interesting due to
the fact that one of the objects, namely $\Bcc$, is coisotropic.
As a result, the space of morphisms \eqref{hmath} is not ``local''
(in a sense that it does not localize to a set of points in $Y$)
and, according to \cite{GW}, is the Hilbert space obtained by
quantizing $(M,\omega)$.
Put differently, the results of \cite{AZ,GW} can be interpreted as
a statement that the space of morphisms between two objects,
at least one of which is coisotropic, is closely related to
quantization.\footnote{To be more precise, for this one needs a little bit more:
the restriction of the curvature of the Chan-Paton bundle of $\Bcc$
to the subspace of $Y$ where $\Bcc$ and $\CB'$ have common support
should be non-trivial. In the special case when the restriction is trivial
the space of morphisms \eqref{hmath} is still very interesting and leads
to a theory of $D$-modules (as opposed to quantization), see~\cite{KW}.}

More generally, the study of coisotropic branes and their role
in the construction of the Fukaya category is an outstanding interesting problem.
Although we will not try to solve it in the present paper, we will be able
to gain some insights by using mirror symmetry.

The computation of the space of morphisms \eqref{hmath}
can be simplified if the space $Y$ happens to admit additional structures.
For example, if $Y$ is hyper-K\"ahler then it is often instructive to
look at $\Bcc$ and $\CB'$ from the vantage point of all three complex
structures, $I$, $J$, and $K = IJ$, as well as the corresponding
symplectic structures $\omega_I$, $\omega_J$, and $\omega_K$.
Even though originally we were interested in $\Bcc$ and $\CB'$
as objects in the Fukaya category 
often they can be defined as half-BPS boundary conditions in
the $\CN=(4,4)$ sigma-model of $Y$, which means that they are
also $A$-branes for some other $A$-model of $Y$, and $B$-branes for a certain $B$-model of $Y$.
In particular, the latter implies that \eqref{hmath} can be also
computed in the $B$-model of $Y$:
\be
\CH \; = \; \Ext^*_{Y} (\Bcc,\CB') \,.
\label{hinbi}
\ee

Another example of a useful structure is a Calabi-Yau structure.
In such case, if $Y$ admits a Calabi-Yau metric, one can approach
the computation of \eqref{hmath} in the mirror $B$-model:
\be
\CH \; = \; \Ext^*_{\tilde Y} (\tBcc,\tCB') \,,
\ee
where $\tBcc$ is the mirror of $\Bcc$, and $\tCB'$ is the mirror of $\CB'$.
As we explain below,
the hyper-K\"ahler structure on $Y$ and mirror symmetry can both be very
useful tools in understanding quantization via categories of $A$- and $B$-branes.
However, combining these tools together can double their power!\\

We start our discussion
in the next section with a friendly introduction to the quantization problem.
Our goal is to explain why this problem is interesting and why it is hard.
Along the way, we often illustrate the general ideas and key concepts with
concrete and (hopefully) simple examples, many of which have applications
to representation theory and gauge theory.
After recalling the $A$-model approach to quantization in section \ref{sec:amodel},
we reformulate the problem in the mirror $B$-model and illustrate it in a number
of examples in section \ref{sec:bmodel}.

One of our examples is so rich and important that it deserves a separate section.
Thus, in section \ref{sec:cstheory} we apply the mirror approach to quantization
of Chern-Simons gauge theory, where the classical phase space~$M$ is the moduli space,
$\M_{{\rm flat}} (G,C)$, of flat connections on a Riemann surface~$C$.
One interesting feature of this example is that, for a compact Lie group~$G$,
the coisotropic brane $\Bcc$ is defined only for a discrete set of symplectic structures on~$Y$,
indexed by an integer number $k$ called the ``level.''
Quantization of $M$ leads to a finite-dimensional Hilbert space~$\CH$,
whose dimension is given by the celebrated Verlinde formula \cite{Verlinde}.
In general, the Verlinde formula has the following form:
\be
\dim \CH \; = \; a_n k^n + a_{n-1} k^{n-1} + \ldots + a_1 k + a_0 \,,
\label{verlindegeneral}
\ee
where $a_i$ are rational numbers.
One novelty of our approach is that it offers an interpretation of
the coefficients $a_i$ in terms of branes on moduli spaces of Higgs bundles.
The coefficients $a_n, a_{n-1}, \ldots$ determine the asymptotic behavior
of this polynomial in the ``classical'' limit $\hbar = \frac{1}{k} \to 0$.
Similarly, 
the coefficients $a_0, a_1, \ldots$ determine the behavior
of the polynomial \eqref{verlindegeneral} in the opposite, ``very quantum'' regime $\hbar = \frac{1}{k} \to \infty$
which, as we explain in section \ref{sec:cstheory},
corresponds to the classical limit ${}^L{\hbar} = - \frac{1}{\hbar} \to 0$
of the mirror theory based on the Langlands dual group $\LG$.

As a result, the coefficients $a_n, a_{n-1}, \ldots$ have a simple interpretation
(in terms of classical geometry of $Y$) and are easier to determine in the $A$-model
based on the moduli space of Higgs bundles with the structure group $G$.
On the other hand, the coefficients $a_0, a_1, \ldots$ have a simpler
interpretation and are easier to determine in the dual $B$-model, based on
the moduli space of Higgs bundles for the Langlands dual group $\LG$.
In section \ref{sec:punctorus}, we present a derivation of the Verlinde formula
using this approach in a concrete example.


\section{Quantization is an art}
\label{sec:qart}

\hfill{\vbox{\hbox{\it Very interesting theory --- it makes no sense at all.}}}

\hfill{\vbox{\hbox{Groucho Marx (about Quantum Mechanics)}}}

\medskip


The basic problem of quantization begins with a symplectic manifold $M$,
called a classical ``phase space,'' equipped with a symplectic form $\omega$.
By quantizing $(M,\omega)$ one can mean a number of different things,
but usually one is asking for a machinery that allows
to turn the following ``classical'' objects into their ``quantum'' analogs:
\medskip
\be
\begin{array}{ccc}
\label{quantizationmap}
 (M,\omega) & \leadsto & \CH ~(=\text{Hilbert space}) \\
  && \raisebox{-.1cm}{\includegraphics[width=.5cm]{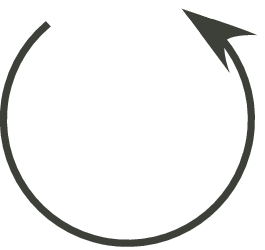}} \\
 \text{alg. of functions on $M$} & \leadsto & \text{~~alg. $\CA_{\hbar}$ of operators on $\CH$} \\
 f & \mapsto & \CO_f :\CH\to\CH \vspace{3mm}\\
 \text{Lagrangian submanifolds} & \leadsto & \text{vectors} \\
  L \subset M & \mapsto & \psi \in \CH \vspace{3mm}\\
 \text{symplectomorphisms} & \leadsto & \text{automorphisms} \\
  \text{of}~ M && \text{of}~ \CA_{\hbar} \vspace{1mm}
\end{array} \ee
There are various interrelations between the classical structures on the left-hand side of
this list, which should be reflected in their quantum counterparts (the right-hand side).
Moreover, depending on specific applications, one can put more items
to this ``wish list''; here we listed only the standard ones.

Since the only input data is $(M,\omega)$ it is not surprising that
all of the items on the left-hand side of \eqref{quantizationmap}
are the standard gadgets in symplectic geometry.
Therefore, quantization can be regarded as a program of constructing
a ``quantum version'' of symplectic geometry.

Another area where the input data is a symplectic manifold is mirror symmetry.
Much like the problem of quantization, it starts with a symplectic manifold
and constructs the Gromov-Witten invariants, the Fuakaya category, and many other
interesting invariants, some of which are even called ``quantum'' ({\it e.g.} quantum cohomology).
Is there any relation between these two problems?

As we explain below, the answer is ``yes'' and the quantization problem
can indeed be reformulated as a certain problem in mirror symmetry,
however, not in the most naive and obvious way.
In particular, the problem of quantizing a symplectic manifold $(M,\omega)$
can be directly related to a problem in the Fukaya category of {\it another}
symplectic manifold, namely a complexification of $(M,\omega)$.

However, before we are ready to review the results of \cite{GW}
and formulate them in terms of mirror symmetry, we need to explain
some of the delicate features of quantization and to introduce important examples.
Since the real dimension of a symplectic manifold is always even,
the simplest non-trivial example is either a 2-sphere, $M = {\bf S}^2$,
or a 2-dimensional plane, $M = \R^2$
(depending on whether we prefer compact or non-compact manifolds).

\example{Quantization of $M = {\bf S}^2$}
One can represent $(M,\omega)$ as a unit sphere in a 3-dimensional space $\R^3$,
\be
x^2 + y^2 + z^2 = 1 \,,
\label{xyzsphere}
\ee
with a symplectic form
\be
\omega = \frac{1}{4\pi \hbar} \frac{dx \wedge dy}{z} \,.
\label{omsphere}
\ee
While it may not be immediately obvious, the 2-form $\omega$
is invariant under the $SO(3)$ symmetry of eq. \eqref{xyzsphere}.
Indeed, using a relation between the Cartesian coordinates $(x,y,z)$
and the spherical coordinates $(r, \theta, \varphi)$,
\begin{eqnarray}
x & = & r \sin \theta \cos \varphi \nonumber \\
y & = & r \sin \theta \sin \varphi \nonumber \\
z & = & r \cos \theta \nonumber
\end{eqnarray}
one can write \eqref{omsphere} as a multiple of the standard volume form on a 2-sphere,
$\omega = \frac{1}{4\pi \hbar} \sin \theta ~d\theta \wedge d \varphi$.
According to textbooks, quantization of $(M,\omega)$ gives a finite-dimensional
Hilbert space $\CH$, such that
\be
\dim \CH = \int_M \omega = \frac{1}{\hbar} \,.
\label{dimhsphere}
\ee
In particular, $\dim \CH$ must be an integer and this shows that
quantization of $(M,\omega)$ is possible only for discrete values
of the parameter $\hbar$:
\be
\hbar^{-1} \; \in \; \Z \,.
\label{hintsphere}
\ee
\endexample

In what follows, we consider a variety of interesting examples
in which symplectic manifolds come from problems in representation theory
on one hand, and from gauge theory and low-dimensional topology, on the other.
These examples provide an excellent laboratory for the quantization problem.

As a necessary preliminary to both groups of examples,
we introduce the following notations that will be used throughout the rest of this paper:

\medskip
\begin{itemize}

\item[$G~$]
$=~$ (simple) compact Lie group,

\item[$G_{\C}$]
$=~$ complexification of $G$,

\item[$G_{\R}$]
$=~$ real form of $G_{\C}$.

\end{itemize}
\medskip

\noindent
In particular, $G_{\R}$ may be equal to $G$.
For concreteness, one can keep in mind a simple example of $G=SU(2)$,
for which $G_{\C} = SL(2,\C)$ and $G_{\R}$ can be either a compact real form $SU(2)$
(equal to $G$) or a split real form $SL(2,\R)$.\\

Now, we can proceed to some interesting examples of symplectic manifolds.
As we mentioned earlier, a large supply comes from representation theory.
Let $\CO_{\R} (\lambda) = G_{\R} \cdot \lambda$ be a coadjoint orbit through an element $\lambda \in \frak g_{\R}^*$,
where $\frak g_{\R} = {\rm Lie} (G_{\R})$ and $\frak g_{\R}^*$ denotes its dual.
To avoid cluttering, we often write $\CO_{\R} (\lambda)$ simply as $\CO_{\R}$.
Then, any such coadjoint orbit is an example of a symplectic manifold \cite{Kostant}.
Indeed, $M=\CO_{\R}$ comes equipped with
the Kostant-Kirillov-Souriau Poisson structure / symplectic structure
that can be written explicitly
\be
\pi = \omega^{-1} = f^{ij}_k X^k \p_i \wedge \p_j
\label{kkpoisson}
\ee
in terms of the structure constants $f^{ij}_k$ of $\frak g_{\R}$.

\example{$G_{\R}=SU(2)$}
In this case, the stabilizer of a generic element $\lambda \in \frak g_{\R}^*$ is a one-dimensional
(abelian) subgroup of $G_{\R} = SU(2)$, so that $\CO_{\R} = SU(2)/U(1) \cong {\bf S}^2$
is simply a 2-sphere, as in our previous example.
Moreover, since the structure constants are given by the totally antisymmetric symbol,
$f^{ijk} = \epsilon^{ijk}$, the Kirillov-Kostant symplectic form \eqref{kkpoisson}
coincides with the one written in \eqref{omsphere} if we identify the three-dimensional space $\R^3$
parametrized by $(x,y,z)$ with (the dual of) the Lie algebra $\frak g_{\R} = \frak{su} (2)$.
Therefore, these two examples are in fact identical.
\endexample

Since the classical phase space $M = \CO_{\R}$ enjoys the action of
the symmetry group $G_{\R}$, this property should be reflected in its quantum counterpart.
Namely, the Hilbert space $\CH$ obtained from quantization of $(M,\omega)$ should carry
a unitary representation of the group $G_{\R}$. This is the basic idea of the orbit method.

However, there have always been some puzzles with this approach to representations
of real groups, which can serve us as important lessons for understanding
the quantization problem:

\begin{itemize}

\item
there exist unitary representations that don't appear to correspond to orbits;

\item
conversely, there are real orbits that don't seem to correspond to unitary representations.

\end{itemize}

\noindent
An example of the first problem occurs even in the basic case
of the real group $G_{\R} = SL(2,\R)$ and the complementary series representations.
To illustrate the second phenomenon, one can take $G_{\R}$ to be
a real group of Cartan type $B_N$, {\it i.e.} $G_{\R} = SO(p,q)$ with $p+q=2N+1$.
The minimal orbit $\CO_{{\rm min}}$ of $B_N$ is a nice symplectic manifold
of (real) dimension $4N-4$, for any values of $p$ and $q$.
On the other hand, the corresponding representation of $SO(p,q)$
exists only if $p \le 3$ or $q \le 3$, and does {\it not} exist if $p,q \ge 4$.
This curious observation \cite{Vogan} follows from a rather lengthy algebra
and cries out for a simple geometric interpretation!

In other words, it would be desirable to have a set of simple topological and/or geometric criteria
that, starting with a symplectic manifold $(M,\omega)$, would tell us beforehand whether
it should be quantizable or not.
Such criteria naturally emerge in the brane quantization approach \cite{GW}
where both of the aforementioned issues can be resolved at the cost of
of replacing classical geometric objects (namely, coadjoint orbits)
with their ``stringy'' analogs (branes).
In particular, in the case of $B_N$ one finds that,
while the minimal orbit exists for any values of $p$ and $q$,
the corresponding brane exists only if $p \le 3$ or $q \le 3$.
(In general, the condition is that the second Stieffel-Whitney class
$w_2(M) \in H^2(M;\Z_2)$ must be
a mod 2 reduction of a torsion class in the integral cohomology of $M$.)\\

Our second class of examples (in fact, also related to representation theory)
comes from gauge theory and low-dimensional topology.
Namely, let us consider Chern-Simons gauge theory
with a real gauge group $G_{\R}$ (that may be compact or not).
The key ingredients in any gauge theory include a gauge connection $A$
and the partial differential equations (PDEs) that it obeys.
In the context of Chern-Simons theory, the relevant equations
are the flatness equations and, according to Atiyah and Bott \cite{AB},
the moduli space of flat connections on a compact oriented 2-manifold
is a finite-dimensional symplectic manifold (possibly singular).

Specifically, let $A$ be a connection on a $G_{\R}$ bundle $E \to C$
over a genus-$g$ Riemann surface $C$.
Then, the moduli space of flat connections on $C$,
$M = \M_{{\rm flat}} (G_{\R} ,C)$, is the space of homomorphisms
$\pi_1 (C) \to G_{\R}$ modulo gauge transformations ({\it i.e.} modulo conjugation).
In order to get a better idea of what this space looks like,
we can describe it more concretely by introducing $G_{\R}$-valued holonomies,
$A_i$, $B_j$, $i,j = 1, \ldots, g$ of the gauge connection over a complete basis
of $A$-cycles and $B$-cycles.
Then, the space $M = \M_{{\rm flat}} (G_{\R} ,C)$ can be viewed as a space
of solutions to the equation
\be
A_1 B_1 A_1^{-1} B_1^{-1} ~\ldots~ A_g B_g A_g^{-1} B_g^{-1} = {\bf 1}
\label{abholonomies}
\ee
modulo conjugation by $G_{\R}$.
In total, the group elements $A_i$ and $B_j$ contain $2g \dim G_{\R}$ real parameters,
so that generically, for $g>1$, after imposing the equation \eqref{abholonomies}
and dividing by conjugation we obtain a space of real dimension $\dim M = 2 (g-1) \dim G_{\R}$.

The space $M = \M_{{\rm flat}} (G_{\R} ,C)$ comes equipped with a natural symplectic form
\be
\omega = \frac{1}{4\pi^2 \hbar} \int_C \Tr \delta A \wedge \delta A \,,
\label{omcs}
\ee
where the parameter $k:=\frac{1}{\hbar}$ is called the ``level.''
What does one find in quantizing $(M,\omega)$?
In particular, what is the Hilbert space $\CH$?
What is the dimension of $\CH$?

The answer to these questions turns out to be surprisingly rich,
and depends in a crucial way on the choice of $G_{\R}$.
If $G_{\R} = G$ is compact, the space $(M,\omega)$ is quantizable
only for integer values of the level,
\be
k = \frac{1}{\hbar} \; \in \; \Z
\label{khinteger}
\ee
In this case, the corresponding Hilbert space $\CH$ is finite-dimensional,
and $\dim \CH$ is a polynomial in $k$, whose leading coefficient
equals the volume of $M$ with respect to the symplectic form $\omega$,
{\it cf.} \eqref{dimhsphere},
\be
\dim \CH = \int_M \frac{\omega^n}{n!} + \ldots \,,
\label{hdimvolm}
\ee
where $\dim M = 2n$.
Specifically, $\CH$ is the space of conformal blocks in the WZW model at level $k$,
and the dimension of $\CH$ is given by the celebrated Verlinde
formula \cite{Verlinde} (see \cite{Beauville} for a nice review).
This is only the beginning of a very beautiful story that leads
to the Witten-Reshetikhin-Turaev invariants of knots and 3-manifolds.

\example{$G=SU(N)$}
The dimension of $\CH$ is given by the following explicit formula:
\be
\dim \CH = \left( \frac{N}{k+N} \right)^{g}
~\sum_{{S \amalg T = [1,k+N] \atop |S| = N}}
~\prod_{{s \in S \atop t \in T}} \vert 2 \sin \pi \frac{s-t}{k+N} \vert^{g-1} \,.
\label{hdimforsun}
\ee
Notice, from this formula it is completely non-obvious that $\dim \CH$
is a polynomial in $k$, {\it e.g.} for $G=SU(2)$ and $g=2$ it gives:
\be
\dim \CH = \frac{1}{6} k^3 + k^2 + \frac{11}{6} k + 1 \,.
\label{dimhgenus2}
\ee
Here, the leading term
equals ${\rm Vol} (M) = {\rm Vol} (\cp^3) = \frac{1}{6} k^3$,
in agreement with \eqref{hdimvolm} and the well-known fact $M \cong \cp^3$ for $G=SU(2)$ and $g=2$.
\endexample

The story is very different if $G_{\R}$ is non-compact.
In this case, the Hilbert space $\CH$ is infinite-dimensional,
and much less is known about the corresponding quantum group invariants.
In particular, the analogs of the Witten-Reshetikhin-Turaev invariants
are still waiting to be discovered.

The two general classes of examples considered here --- based on coadjoint orbits
and moduli spaces of flat connections --- are actually much closer related than one might think.
Indeed, a coadjoint orbit $\CO_{\R}$ (more precisely, the corresponding conjugacy
class $\frak C_{\R} \subset G_{\R}$) naturally appears as a ``local model''
for $\M_{{\rm flat}} (G_{\R} ,C)$ if we take $C$ to be a punctured disc, see {\it e.g.} \cite{Ramified}.
In fact, these two classes of examples can be naturally combined in a larger
family by picking a set of ``marked points'' $p_i$, $i=1,\ldots,h$ on the Riemann surface $C$
and requiring the gauge field $A$ to have certain singularities at the points $p_i$.
Equivalently, one can remove the points $p_i$ and study Riemann surfaces with
punctures (or boundary components).

For ease of exposition,
let us consider a Riemann surface with only one puncture $p \in C$,
around which the gauge field has a holonomy
\be
V = {\rm Hol}_{p} (A) \; \in \; \frak C_{\R} \,,
\label{aholonomy}
\ee
that takes values in a prescribed conjugacy class $\frak C_{\R} \subset G_{\R}$.
In this way, associating a conjugacy class to a puncture,
we obtain the moduli space $M = \M_{{\rm flat}} (G_{\R} ,C; \frak C_{\R})$
of flat connections on $C \setminus p$.
As in \eqref{abholonomies}, this moduli space can be described rather explicitly
as a space of solutions to the equation
\be
A_1 B_1 A_1^{-1} B_1^{-1} ~\ldots~ A_g B_g A_g^{-1} B_g^{-1} = V \,,
\label{parabhol}
\ee
modulo conjugation by $G_{\R}$.

The moduli space $M = \M_{{\rm flat}} (G_{\R} ,C; \frak C_{\R})$
is a symplectic manifold, with the symplectic form $\omega$
given by the general formula \eqref{omcs}.
At least when $V$ is sufficiently close to ${\bf 1}$,
it has the structure of a symplectic fibration
\be
\begin{matrix}
\frak C_{\R} & \to & \M_{{\rm flat}} (G_{\R} ,C; \frak C_{\R}) \\
 & & \downarrow \! \imath \\
 & & ~\M_{{\rm flat}} (G_{\R} ,C)
\end{matrix}
\label{mparabfibr}
\ee
Furthermore, the symplectic form on $\M_{{\rm flat}} (G_{\R} ,C; \frak C_{\R})$ is
\be
\omega = \imath^* \omega_{\M} + \omega_{\frak C} \,,
\label{omparabfibr}
\ee
where $\omega_{\M}$ is the symplectic form on $\M_{{\rm flat}} (G_{\R} ,C)$
and $\omega_{\frak C}$ restricts to the Kostant-Kirillov-Souriau symplectic
form \eqref{kkpoisson} on each fiber of the symplectic fibration~\eqref{mparabfibr}.

\example{$G=SU(2)$}
Unitary irreducible representations\footnote{The representations
with even $\lambda$ are also $SO(3)$ representations.}
of $G=SU(2)$ can be labeled either by the highest weight $\lambda \in \Z_{\ge 0}$
or, equivalently, by the dimension $d = \lambda + 1$.
In the physics literature, a representation $R_{\lambda} = S^{\lambda} \C^2$
is often called the spin-$j$ representation, where $j=\lambda/2$.
As in \eqref{aholonomy},
we can associate a representation $R_{\lambda}$ to a marked point $p \in C$
by making a puncture, such that on a small loop around $p$ the gauge field
has a holonomy conjugate to:
\be
V \; = \; \exp 2 \pi i
\begin{pmatrix}
\alpha & 0 \\ 0 & -\alpha
\end{pmatrix} \,,
\label{su2holonomy}
\ee
with $\alpha = \frac{\lambda}{2k}$.
Then, for a Riemann surface of genus $g$ with $h$ punctures, the Verlinde formula is
\be
\dim \CH \; = \; \left( \frac{k+2}{2} \right)^{g-1}~
\sum_{j=1}^{k+1} \left( \sin \frac{\pi j}{k+2} \right)^{2-2g-h}
\prod_{i = 1}^h \sin \frac{\pi j (\lambda_i + 1)}{k+2} \,.
\label{verlindeparabolic}
\ee
Believe it or not, this is an integer!
\endexample

Now, once we introduced a good supply of interesting symplectic manifolds,
we shall return to our original problem of quantization of $(M,\omega)$.
The quantization problem can be approached in many different ways.
Each approach has its advantages and disadvantages,
but in the end all methods are expected to yield the same result.
Below we give a brief overview of various methods, quickly specializing
to the brane quantization approach that will be used in the rest of this paper.


\subsection{Geometric quantization}

In geometric quantization \cite{Kostant2,Souriau}, in order to produce the desired items
on the right-hand side of \eqref{quantizationmap} one first needs to introduce
some extra data that is not supplied with the symplectic manifold $(M,\omega)$.
Then, of course, one needs to show that the result is, in a suitable sense,
independent on these auxiliary choices.
(This last step turns out to be the most difficult one almost in every
approach to quantization.)

The first piece of extra data --- which one needs to introduce not only in
geometric quantization, but more or less in any approach to quantization ---
is a choice of line bundle, $\CL \to M$, called the ``prequantum line bundle''
with a unitary connection of curvature $\omega$.
Note, a prequantum line bundle $\CL \to M$ only exists for
\be
[\omega] \in H^2(M;\Z) \,,
\label{omegaquant}
\ee
which can lead to a quantization of $\hbar^{-1}$
({\it i.e.} a restriction of $\hbar$ to a discrete set of values in $\C^*$).
In fact, we already saw this phenomenon
in our examples in \eqref{hintsphere} and \eqref{khinteger}.

The second choice of extra data is more delicate: it is a choice of polarization
that, on local charts, corresponds to representing $M$ as a cotangent bundle $T^* U$.
It is this second step where geometric quantization faces serious difficulties.
Even if one can locally represent $M \simeq T^* U$ in every chart,
such choices may not agree globally.
Moreover, showing that the answer is independent of such choices becomes
a rather difficult task.


\subsection{Deformation quantization}

Deformation quantization involves no auxiliary choices \cite{BFFLS}.
However, it is {\it not} a quantization in the sense of \eqref{quantizationmap}.
Indeed, it does not construct the Hilbert space $\CH$ and gives
only a formal deformation of the ring of functions on $M$.
In deformation quantization, there is no quantization condition on
the parameter $\hbar$.


\subsection{Brane quantization}
\label{sec:amodel}

As in other quantization methods, the approach of \cite{GW} starts with
a number of auxiliary choices that we summarize below:

\medskip
\begin{itemize}

\item
$Y=$ complexification of $M$, {\it i.e.} a complex manifold equipped
with a complex structure that we shall call $J$
and an antiholomorphic involution $\tau : Y \to Y$,
such that $\tau^* J = -J$ and $M$ is contained\footnote{After
embedding the quantization problem in the $A$-model of $Y$,
it is natural to replace the latter condition with a slightly more
general one, $\tau: M \to M$. Among other things, this generalization
turns out to be important for finding ``missing'' coadjoint orbits corresponding
to the complementary series representations \cite{GW}.} in the fixed point set of~$\tau$,

\item
$\Omega =$ 
(non-degenerate) holomorphic 2-form, such that
$\tau^* \Omega = \bar \Omega$ and
$$
\Omega \vert_M = \omega \,,
$$

\item
$\CL \to Y$ unitary line bundle (extending the ``prequantum line bundle'' $\CL \to M$)
with a connection of curvature ${\rm Re}\, \Omega$.

\end{itemize}

\medskip
\noindent
Of course, these data need to be consistent.
For example, we need to ask for $\tau$ to lift to an action on $\CL \to Y$,
such that $\tau \vert_M = {\rm id}$, {\it etc.}

To summarize, the basic idea is to pass from the original symplectic
manifold $(M, \omega)$ and the prequantum line bundle $\CL$
(that we often regard as a part of the initial data) to the complexification $(Y, \Omega)$ and $\CL$.
Then, the problem of quantizing $(M,\omega)$ can be formulated
as a problem in the $A$-model / Fukaya category of $Y$ with symplectic structure
\be
\omega_Y = - {\rm Im}\, \Omega \,.
\label{omydef}
\ee
Note, the symplectic structure $\omega_Y$ is not a part of the original data,
and appears only after we complexify the original phase space $M$.

Before we explain how all the desired items on the right-hand side of \eqref{quantizationmap}
can be produced in the $A$-model of $(Y,\omega_Y)$, it is important to emphasize that in this
approach to quantization the focus shifts from $M$ to $Y$, so that $Y$ takes the center of the stage.
Then, from the vantage point of $Y$ it may be natural to consider close cousins
of the original quantization problem suggested by the analysis of the $A$-model / Fukaya category of $Y$.
For example, one can reduce the list of the auxiliary choices (see above)
by omitting the involution $\tau$, which is needed only for unitarity.
If one does not require this extra structure (namely, the Hermitian inner product on $\CH$),
then it suffices to introduce the complex symplectic manifold $(Y,\Omega)$
with a line bundle $\CL \to Y$, and no involution $\tau$.
From these data alone one can construct a space $\CH$ and an algebra $\CA_{\hbar}$ that acts on it.
Later we consider examples of such situations related to representations theory.

We shall illustrate this approach to ``quantization via complexification''
in a variety of examples introduced earlier in this section;
in particular, we apply it to $M=\CO_{\R}$ and $M = \M_{{\rm flat}} (G_{\R} ,C)$.
Although these examples have a very different flavor and come from
completely different areas of physics and mathematics, they are closely
related to representation theory of real groups and, at the most basic level,
the complexification of $M$ can be understood as passing from a real group $G_{\R}$
to its complexification $G_{\C}$.

\example{Quantization of $M=\CO_{\R}$}
A coadjoint orbit $\CO_{\R}$ of a real group $G_{\R}$ admits an obvious complexification,
namely a complex coadjoint orbit of $G_{\C}$:
\be
Y = \CO_{\C} \,.
\label{yorbitcplx}
\ee
For example, the real coadjoint orbit \eqref{xyzsphere} of $G_{\R} = SU(2)$
has a complexification $Y=\CO_{\C}$ described by the same equation $x^2 + y^2 + z^2 = 1$,
where $x$, $y$, and $z$ are now complex variables.
Moreover, eq. \eqref{omsphere} written in terms of $(x,y,z) \in \C^3$
defines a holomorphic symplectic form $\Omega$ on $Y$.
\endexample

Similarly, there is an obvious complexification of the moduli space
of flat connections, $\M_{{\rm flat}} (G_{\R} ,C)$.

\example{Quantization of $M = \M_{{\rm flat}} (G_{\R} ,C)$}
This symplectic manifold $M$ admits an obvious complexification:
\be
Y = \M_{{\rm flat}} (G_{\C} ,C) \,,
\label{yflatcplx}
\ee
the moduli space of flat $G_{\C}$ connections on $C$.
Much like $M$ itself, the space $Y$ can be explicitly described
as a space of $G_{\C}$-valued holonomies $A_i$ and $B_j$ that
satisfy \eqref{abholonomies}, modulo conjugation by $G_{\C}$.
It comes equipped with a holomorphic symplectic form $\Omega$
which, in terms of the $\frak g_{\C}$-valued gauge connection,
has the familiar form \eqref{omcs}.
\endexample

\noindent
In these examples, it is easy to verify that the holomorphic symplectic
form $\Omega$ restricts to $\omega$ on $M \subset Y$.\\

Now let us return to the quantization problem of $(M,\omega)$
and explain how the desired items on the right-hand side of \eqref{quantizationmap}
can be produced in the approach of~\cite{GW}.
The Hilbert space $\CH$ is constructed as the space
of morphisms (space of open strings),
\be
\CH \; = \; {\rm Hom} (\Bcc,\CB') \,,
\label{hbbcc}
\ee
where $\CB_{cc}$ and $\CB'$ are objects (branes)
of the Fukaya category of $(Y,\omega_Y)$.

In general, typical objects of the Fukaya category of $(Y,\omega_Y)$
are Lagrangian submanifolds of $Y$ equipped with flat unitary vector bundles
and, in our setup, $\CB'$ is exactly such an object.
Specifically, we define $\CB'$ to be an $A$-brane supported on $M \subset Y$.
Indeed, according to our definitions,
\be
\omega_Y \vert_M = 0 \,,
\ee
so that $M$ is a Lagrangian submanifold of $Y$ with respect to $\omega_Y$.

Less familiar examples of $A$-branes (objects of the Fukaya category)
are coisotropic submanifolds of $(Y,\omega_Y)$ equipped with non-flat
vector bundles (a.k.a. Chan-Paton bundles) with unitary connection
that obeys certain conditions \cite{KO}.
In the simplest case of rank-1 coisotropic objects supported on all of $Y$,
the condition on the curvature 2-form $F$ is
\be
(\omega_Y^{-1} F)^2 = -1 \,.
\label{fcois}
\ee
In our approach to quantization, $\CB_{cc}$ is an example of such object,
namely the so-called {\it canonical coisotropic} brane
associated to a complexification of $(M,\omega)$ in a canonical way \cite{GW}.

To summarize, after we choose a complexification of $(M,\omega)$
and the extension of the prequantum line bundle $\CL$,
we can define two canonical objects in the Fukaya category of $(Y, \omega_Y)$:

\begin{framed}
\begin{itemize}

\item[$\CB' \; = \; $]
Lagrangian $A$-brane supported on $M \subset Y$\\

\item[$\Bcc \; = \;$]
coisotropic $A$-brane supported on $Y$ and endowed with a unitary line bundle $\CL$
with a connection of curvature
$$
F = {\rm Re}\, \Omega
$$

\end{itemize}
\end{framed}

\noindent
In particular, it is easy to verify that, with our definition of $\omega_Y$,
the curvature 2-form $F$ indeed obeys the required condition \eqref{fcois}.

Given two objects $\CB'$ and $\Bcc$, it is natural to consider the spaces
of morphisms (spaces of open strings) in the $A$-model of $(Y,\omega_Y)$.
As we already stated in \eqref{hbbcc}, the space of $(\Bcc,\CB')$ strings
gives the Hilbert space $\CH$ associated with the quantization of $(M,\omega)$.
Just like in other quantization methods, one needs to show that it is
independent on the auxiliary choices (which, among other things, involve
the choice of complex structure $J$ on $Y$), {\it i.e.} to construct a flat
connection on the $\CH$-bundle over the space of such choices.
In a closely related context, this problem has been studied in the mathematical physics
literature \cite{CecottiV,Dubrovin},
and leads to a beautiful story that involves integrable systems and $tt^*$ equations.

\example{Quantization of $M = T^2$}
In this problem, $M = T^2$ admits an obvious complexification,
\be
Y \; \cong \; \C^* \times \C^* \,,
\label{yccstar}
\ee
and the resulting Hilbert space $\CH$ should not depend, among other things,
on the choice of complex structure on $M = T^2$. If we denote by $t \in \CT$
the corresponding complex structure parameter,
then the states in $\CH$ are simply theta functions of order $k$,
\be
\vartheta_r (z; t) = \sum_{n=- \infty}^{\infty}
\exp \left( \frac{\pi i t}{k} ( kn + r )^2 + 2 \pi i ( kn + r) z \right)
\quad\quad r \in \Z / k \Z \,,
\label{jactheta}
\ee
where $k: = \dim \CH = \frac{1}{\hbar}$, {\it cf.} \eqref{khinteger}.
It is easy to see from \eqref{jactheta} that $\vartheta_r (z; t)$
are quasi-periodic,
$$
\vartheta_r (z + a + b t ; t)
\; = \; \exp \left( - \pi i k b^2 t - 2 \pi i k bz \right) \vartheta_r (z ; t)
\quad\quad\quad a,b \in \Z
$$
and obey the heat equation
\be
\left( {\p \over \p t} - {\hbar \over 4 \pi i} \; {\p^2 \over \p z^2} \right)
\vartheta_r (z; t) = 0 \,,
\ee
which gives a connection on a bundle $\CH \to \CT$.
This example will be approached from a different viewpoint in section \ref{sec:torus}.
\endexample

Furthermore, in brane quantization
the involution $\tau$ leads to a Hermitian inner product on $\CH$.
It is not necessarily positive definite; a necessary condition is that
$\tau$ fixes $M$ pointwise outside of a compact support.\footnote{Indeed,
in the classical limit the norm of a state $\psi \in \CH$ is roughly
$$
\langle \psi , \psi \rangle = \int_M \bar \psi (\tau x) \psi (x) \,.
$$
It is positive definite only if $\tau$ fixes $M$ pointwise.}
This slight generalization of the condition that $M$ belongs to the fixed point set of $\tau$
is important {\it e.g.} for constructing the complementary series representations,
where $\tau$ acts non-trivially on $M$.

The space of $(\Bcc,\Bcc)$ strings, on the other hand, gives an associative but non-commutative
algebra,
\be
\CA_{\hbar} \; = \; {\rm Hom} (\Bcc,\Bcc) \,.
\label{abccbcc}
\ee
Note, this algebra depends only on $(Y,\Omega)$ and not on $M$.
(In our examples, it means that the same algebra $\CA_{\hbar}$
acts on Hilbert spaces obtained in quantization of $M = \M_{{\rm flat}} (G_{\R} ,C)$
for different real forms $G_{\R}$ of $G_{\C}$, and similarly for $M=\CO_{\R}$.)
In fact, we can think of the algebra $\CA_{\hbar}$
as arising from the deformation quantization of~$Y$.

The path integral of the quantum mechanics on $M$ also has
an elegant realization in the $A$-model approach, see \cite{newlook} for details.


\subsection{When does quantization exist?}

While highly desirable, a complete set of geometric criteria that determine
which symplectic manifolds are quantizable (and which are not) is not known at present.
Most likely, such criteria should include the condition \eqref{omegaquant} that controls
the existence of the prequantum line bundle $\CL$ (and sometimes leads to a quantization of $\hbar$).
However, this condition alone is clearly not enough, and --- even as some of our examples suggest ---
there should be further criteria which determine whether $(M,\omega)$ is quantizable or not.

{}From the viewpoint of brane quantization, $(M,\omega)$ is expected to be quantizable
whenever it admits a complexification, such that $(Y,\omega_Y)$ has a ``good'' $A$-model / Fukaya category.
A precise necessary condition for this is not known
(in part, since the present understanding of the Fukaya category is incomplete).
A sufficient condition, though, is that $(Y, \omega_Y)$ admits a complete Calabi-Yau metric $g$,
for which $\omega_Y$ is a K\"ahler form, {\it i.e.} $K = g^{-1} \omega_Y$ is an integrable complex structure.
One indication that such criteria are on a right track is that,
in deformation quantization, one encounters similar conditions
that tell us whether $\CA_{\hbar}$ is an actual deformation of
the algebra of holomorphic functions on $(Y,\Omega)$,
with a complex parameter $\hbar$ (not just a formal variable).

\example{Quantization of $M = {\bf S}^2$}
In \eqref{xyzsphere} we represented ${\bf S}^2$ as a unit sphere in $\R^3$.
Its complexification,
\be
x^2 + y^2 + z^2 = 1 \,,
\qquad\qquad
(x,y,z) \in \C^3
\label{ehcplxsurf}
\ee
admits a complete Calabi-Yau metric (the Eguchi-Hanson metric)
and a deformation of the ring of functions with a complex parameter $\hbar$ (not just a formal deformation).
Both of these properties fail for a complex surface,
\be
x^4 + y^4 + z^4 = 1 \,,
\qquad\qquad
(x,y,z) \in \C^3
\ee
that can be viewed as an alternative complexification of $M = {\bf S}^2$.
\endexample

Besides the requirement for $(Y, \omega_Y)$ to have a good $A$-model / Fukaya category,
one needs $\CB'$ and $\Bcc$ to exist in order to solve the original quantization
problem, {\it i.e.} to compute the spaces of morphisms \eqref{hbbcc} and \eqref{abccbcc}.
Fortunately, the existence of $\CB'$ and $\Bcc$ can be expressed in terms of concrete geometric criteria,
which can be useful even in other quantization methods.
Specifically, the brane $\CB'$ supported on $M$ exists whenever $M$ admits a flat Spin$^c$ structure,
and the brane $\Bcc$ exists whenever $[{\rm Re} \, \Omega] \in H^2(Y;\Z)$, {\it cf.} \eqref{omegaquant}.


\section{$B$-model approach to quantization}
\label{sec:bmodel}

\hfill{\vbox{\hbox{\it Gott w\"urfelt nicht!}}}

\hfill{\vbox{\hbox{Albert Einstein}}}

\medskip


In section \ref{sec:amodel} we reviewed the $A$-model approach to
quantization \cite{GW}, where to a classical symplectic manifold $(M,\omega)$
one associates a Fukaya category of $A$-branes and
the quantization \eqref{quantizationmap} is achieved
by studying the space of morphisms between two branes $\Bcc$ and $\CB'$.
It is important to emphasize, however, that the Fukaya category in
question is {\it not} that of the original symplectic manifold $(M,\omega)$.
Rather, it is the Fukaya category of $Y$, a complexification of $M$,
considered with a new symplectic form \eqref{omydef} that didn't exist
prior to complexification.

Another area of physics $\&$ mathematics where Fukaya categories are of
major importance is mirror symmetry. In general, mirror symmetry relates
the $A$-model of a symplectic manifold $Y$ to the $B$-model of
a complex manifold $\tilde Y$, called the mirror of $Y$.
In mirror symmetry, however, the Fukaya category and $A$-model
are usually considered to be the `difficult' side of the correspondence,
and it is often convenient to use mirror symmetry
to map the problem to the simpler $B$-model side.

In our present context, this map is described by the homological mirror
symmetry conjecture \cite{Kontsevich} that relates the derived Fukaya category
of $Y$ and the (bounded) derived category of coherent sheaves on $\tilde Y$.
Specifically, the conjecture says that there exists a functor:
\be
\Phi_{{\rm mirror}} ~:~~~ {\rm{\bf Fuk}} (Y) \;\overset{\sim}{\longrightarrow}\; \CD^b (\tilde Y)
\label{mirrormap}
\ee
such that it is the equivalence of triangulated categories.
In the rest of this paper our goal will be to apply this map
to the $A$-model of $(Y,\omega_Y)$ described in section \ref{sec:amodel}
and thereby to reformulate the quantization problem entirely
in terms of the $B$-model.

In particular, mirror symmetry maps our $A$-branes $\Bcc$ and $\CB'$
to the dual $B$-branes:
\begin{eqnarray}
\tCB' & = & \Phi_{{\rm mirror}} (\CB') \label{bbmirrors} \\
\tBcc & = & \Phi_{{\rm mirror}} (\Bcc) \nonumber
\end{eqnarray}
whose geometry we wish to explore.
Furthermore, mirror symmetry provides a dual description of
the Hilbert space $\CH$ and the algebra of quantum operators $\CA_{\hbar}$
in terms of $\Ext$-groups of the dual objects $\tCB'$ and $\tBcc$.
For example, it identifies the space of morphisms \eqref{hbbcc}
with
\be
\CH \; = \; \Ext^*_{\tilde Y} (\tBcc,\tCB') \,,
\label{mirrorhbbcc}
\ee
which can be analyzed using the standard tools of algebraic geometry.
Thus, the Euler characteristic of \eqref{mirrorhbbcc} can be easily
computed in the $B$-model with the help of the Grothendieck-Riemann-Roch theorem:
\be
\sum_{k} (-1)^{k} \dim \Ext^k_{\tilde Y} (\tBcc,\tCB')
= \int_{\tilde Y} \ch (\tBcc)^* \wedge \ch (\tCB') \wedge {\rm Td} (\tilde Y) \,,
\label{GRRtheorem}
\ee
where $\ch (\tCB')$ (resp. $\ch (\tBcc)$) is the Chern character of
$\tCB'$ (resp. $\tBcc$), ${\rm Td} (\tilde Y)$ is the Todd class of $\tilde Y$,
and $\omega^*$ denotes $(-1)^{p+1} \omega$ for any $2p$-form $\omega$.
In applications, we will often use \eqref{GRRtheorem} to compute
the dimension of the Hilbert space $\CH$ (when $\dim \CH < \infty$).\\

What are the mirror objects $\tCB'$ and $\tBcc$?
Does the mirror of the canonical coisotropic $A$-brane $\Bcc$
admit a `canonical' definition in
$\CD^b (\tilde Y)$ ({\it i.e.} in the $B$-model of $\tilde Y$)?
What is the role of $\hbar$ in the $B$-model of $\tilde Y$?
In order to answer these and other questions about the $B$-model
approach to quantization of $(M,\omega)$, it is useful to have
a good geometric description of the mirror transform \eqref{mirrormap}.
One such description was proposed in 1996 by Strominger, Yau, and Zaslow \cite{SYZ} (see also \cite{BJSV}),
who argued that mirror Calabi-Yau manifolds $Y$ and $\tilde Y$
should fiber over the same base manifold $\BB$,
\be
\begin{array}{ccc}
\! Y~~~~~~~{} & \; & {}~~~~~~\widetilde{Y} \\
{}~~~~~~~{}_{\pi} \searrow & \; & \!\! \swarrow_{~\tilde \pi}~~~~~~~{} \\
\; & \!\! \BB & \;
\end{array}
\label{syzmirror}
\ee
with generic fibers $\FF_b = \pi^{-1} (b)$ and $\widetilde{\FF}_b = \tilde \pi^{-1} (b)$,
$b \in \BB$, being dual tori, in the sense that
$\FF_b = H^1 (\widetilde{\FF}_b, U(1))$ and $\widetilde{\FF}_b = H^1 (\FF_b, U(1))$.
Moreover, the fibers $\FF_b$ and $\widetilde{\FF}_b$
should be (special) Lagrangian submanifolds\footnote{
We remind that, by definition, a middle-dimensional submanifold $M \subset Y$
is called Lagrangian if the symplectic form $\omega_Y$ vanishes on $M$,
and is {\it special} Lagrangian if, in addition,
the imaginary part of the holomorphic volume form on $Y$ vanishes when restricted to $M$.}
in $Y$ and $\tilde Y$, respectively.

This way of looking at mirror symmetry can be very useful in understanding
how the functor \eqref{mirrormap} acts on the $A$-branes $\CB'$ and $\Bcc$,
which ultimately will lead us to a reformulation of the quantization
problem in the mirror $B$-model.
In particular, as we explain below, the fate of the coisotropic brane $\Bcc$
depends in a crucial way on whether the restriction of the symplectic form
$F = {\rm Re}\, \Omega$ on $Y$ to a generic fiber $\FF$
of the SYZ fibration \eqref{syzmirror} is trivial or not.
When it is trivial, the coisotropic brane $\Bcc$ transforms under
mirror symmetry to a brane $\tBcc$ supported on a middle-dimensional
submanifold of $\tilde Y$, namely on a section of the dual SYZ fibration.
(An example of such situation was considered {\it e.g.} in \cite{KW}.)

In contrast, when $F \vert_{\FF}$ is non-trivial, the story becomes
more interesting and more complicated.
In this case --- which will be our subject here ---
mirror symmetry transforms the coisotropic brane $\Bcc$ into
a $B$-brane $\tBcc \in \CD^b (\tilde Y)$ supported on all of $\tilde Y$.
Furthermore, in general $\tBcc$ is a brane of a fairly high rank.
In fact, using the SYZ picture \eqref{syzmirror} we conclude that
the rank of $\tBcc$ is given by
\be
\boxed{\phantom{\int} \rank (\tBcc) \; = \; {\rm Vol} (\FF)\phantom{\int}}
\label{tbccrank}
\ee
where ${\rm Vol} (\FF)$ is the volume of the SYZ fiber $\FF$
computed with respect to the symplectic form $F = {\rm Re}\, \Omega$ on $Y$,
\be
\rank (\tBcc) \; = \; \int_{\FF} \frac{F^n}{n!} \,,
\label{tbccfrank}
\ee
and $\dim_{\R} \FF = \dim_{\C} Y = 2n$.
(Remember, that in our context $Y$ is always a complex symplectic manifold.)
Notice, the formula \eqref{tbccrank}
also applies to the simpler case where $F \vert_{\FF}$ is trivial.

In what follows, we shall illustrate \eqref{tbccrank} in a variety of concrete examples.
However, there is also a general argument based on \eqref{syzmirror}
that we wish to sketch here since it will be very useful in later applications.
In the $A$-model approach to quantization, $\CB'$ is a Lagrangian brane on $(Y,\omega_Y)$.
Since the fiber of the SYZ fibration \eqref{syzmirror} is Lagrangian
with respect to $\omega_Y = - {\rm Im}\, \Omega$,
we can choose $\CB'$ to be a Lagrangian brane
supported on a generic fiber $\FF_b \subset Y$ and equipped with a unitary flat line bundle.
In this simple warm-up example we know exactly what the dual object $\tCB'$ is.
It is the skyscraper sheaf $\CO_p \in \CD^b (\tilde Y)$ of a point $p \in \tilde Y$,
such that $\tilde \pi (p) = b$. For this reason, $\tCB' = \CO_p$ is often called
a ``zero-brane'' or ``D0-brane'' on $\tilde Y$. Summarizing,
\be
\Phi_{{\rm mirror}} ~:~~~ \CB_{\FF} \; \rightarrow \; \CB_p \,,
\label{syzfp}
\ee
where we used slightly more intuitive notations $\CB_{\FF}$ and $\CB_p$
for this type of $A$-branes and $B$-branes, respectively.

In fact, the mirror pair of branes in \eqref{syzfp} was an important part
of the original motivation in \cite{SYZ} that led to the proposed picture \eqref{syzmirror}.
One way to see that $\CB_{\FF}$ and $\CB_p$ should be mirror to each other
is to consider their self-Homs. For a $B$-brane $\CB_p = \CO_p$ on $\tilde Y$,
we have
\be
\Ext^*_{\tilde Y} (\CB_p, \CB_p) \cong \Lambda^* T_p \tilde Y \cong H^* (T^{2n}, \C) \,.
\ee
As a gradede vector space, it is isomorphic to the Floer cohomology of $\FF \cong T^{2n}$,
which describes the self-Homs of the $A$-brane $\CB_{\FF}$:
\be
HF^* (\CB_{\FF}, \CB_{\FF}) \cong H^* (\FF, \C) \,,
\ee
hence, justifying \eqref{syzfp}.

Now, once we understand the duality \eqref{syzfp} between branes $\CB_{\FF}$ and $\CB_p$,
we can use it to ``probe'' the geometry of $\tBcc$.
Namely, as suggested earlier, we can use $\CB_{\FF}$ for the $A$-brane $\CB'$
(and, hence, $\CB_p$ for the mirror $B$-brane $\tCB'$) to compute the space
of morphisms $\CH$ (= space of open strings) between $\CB_{\FF}$ and $\Bcc$,
just like we did it a moment ago for the brane $\CB_{\FF}$ itself.
According to \eqref{hbbcc}, in the $A$-model of $(Y,\omega_Y)$
the space $\CH = {\rm Hom} (\Bcc,\CB_{\FF})$ is obtained by quantizing
the support of $\CB_{\FF}$,
with the symplectic form $\omega = {\rm Re}\, \Omega \vert_{\FF}$.
Since the support of $\CB_{\FF}$ is an abelian variety $\FF \cong T^{2n}$,
its quantization is well understood
and leads to a space of $\theta$-functions, {\it cf.} \eqref{jactheta},
of dimension\footnote{Notice, eq. \eqref{hdimvolm} is exact in this case.}
\be
\dim \CH = \int_{\FF} \frac{F^n}{n!} \,,
\label{dimhtorus}
\ee
where we used $F = {\rm Re}\, \Omega$.
This gives us the right-hand side of \eqref{tbccfrank}.
On the other hand, calculating the dimension of $\CH$
in the $B$-model of $\tilde Y$ with the help of \eqref{GRRtheorem}
we obtain $\dim \CH = \dim \Ext^*_{\tilde Y} (\tBcc,\CB_p) = \rank (\tBcc)$,
which is precisely the left-hand side of \eqref{tbccfrank}.
This calculation concludes a useful exercise that will also
serve us as a practice example for studying $\CH$ in the $A$-model of $Y$
and in the $B$-model of $\tilde Y$.


\subsection{$(B,B,B)$ branes and hyperholomorphic bundles}
\label{sec:hyperk}

In the $A$-model approach to quantization, the classical phase space $(M,\omega)$
is replaced by a complex symplectic manifold $(Y,\Omega)$ which, by definition,
comes equipped with two symplectic forms that we call
$F = {\rm Re}\, \Omega$ and $\omega_Y = - {\rm Im}\, \Omega$,
and a complex structure $J$ that relates them.

Now we wish to focus on a particularly nice situation
where both symplectic forms $F$ and $\omega_Y$ are K\"ahler with respect
to some complex structures $I$ and $K = IJ$,
so that $Y$ is a hyper-K\"ahler manifold
(this happens {\it e.g.} if $M$ is a K\"ahler manifold).
Then, using the standard notations $\omega_I$, $\omega_J$, $\omega_K$
for the three K\"ahler forms corresponding to the complex structures $I$, $J$, and $K$,
we can write the holomorphic symplectic form $\Omega$ as
\be
i \Omega = \omega_K + i \omega_I \,,
\label{omegahk}
\ee
where, according to the conventions of section \ref{sec:amodel},
\be
F = \omega_I
\qquad , \qquad
\omega_Y = \omega_K \,.
\label{fomhk}
\ee

What about the objects $\Bcc$ and $\CB'$ that play a central role in
the $A$-model approach to quantization? As we already mentioned in the Introduction,
they tend to be automatically compatible with the hyper-K\"ahler structure on $Y$, when it exists.
Namely, defined as half-BPS boundary conditions in the $\CN=(4,4)$ sigma-model of $Y$,
they often preserve supersymmetry in two different $A$-models of $Y$,
with respect to different symplectic forms, say $\omega_J$ and $\omega_K$,
and also in a $B$-model of the third complex structure, $I$.
Following the terminology introduced in \cite{KW},
we call such objects ``branes of type $(B,A,A)$.''

A quick remark on the notation is on order. On a hyper-K\"ahler manifold $Y$
the choice of what we call the complex structures $I$, $J$, and $K$
(and the corresponding K\"ahler forms $\omega_I$, $\omega_J$, and $\omega_K$)
is, of course, entirely up to us.
In fact, $I$, $J$, and $K$ are part of the entire sphere ${\bf S}^2 =  \cp^1$
of complex structures on $Y$,
\be
\CI = a I + b J + c K \,,
\label{cplstrsphere}
\ee
parametrized by $(a,b,c) \in \R^3$ with $a^2 + b^2 + c^2 = 1$.
Therefore, when in a favorable situation we say that $\Bcc$ and $\CB'$
are holomorphic in complex structure $I$ --- which makes them branes of type $(B,A,A)$ ---
this choice is quite random, except that its orientation with respect to
the fiber of the SYZ fibration \eqref{syzmirror} is very important.
Throughout the paper, we adopt the convention that, when $Y$ is hyper-K\"ahler,
the fiber $\FF$ is always holomorphic in complex structure $I$
(and Lagrangian with respect to $\omega_J$ and $\omega_K$).
In other words, the fiber itself is an object (brane) of type $(B,A,A)$.
This makes the complex structure $I$ and, hence, the branes of
type $(B,A,A)$ a bit special among others.

A typical example of a $(B,A,A)$ brane is a middle-dimensional submanifold of $Y$
that is holomorphic in complex structure $I$ and Lagrangian with respect
to both $\omega_J$ and $\omega_K$. In fact, $\CB'$ is a good example of
such an object, when $Y$ is hyper-K\"ahler.

\example{$(B,A,A)$ branes on $Y = \R^4$}
Locally, the geometry of every hyper-K\"ahler manifold looks like
a quaternionic $n$-pane, $\IH^n$. In the simplest case $n=1$,
we may identify a point $(x_0,x_1,x_2,x_3) \in \R^4$ with a quaternion $q \in \IH$:
\be
q = x_0 + {\bf i}\, x_1 + {\bf j}\, x_2 + {\bf k}\, x_3
\ee
where ${\bf i}^2 = {\bf j}^2 = {\bf k}^2 = {\bf i j k} = -1$.
The three complex structures $I$, $J$, $K$ act on $\R^4 \cong \IH$
by left multiplication by ${\bf i}$, ${\bf j}$, ${\bf k}$,
and the corresponding K\"ahler forms are
\be
{\bf i}\, \omega_I + {\bf j}\, \omega_J + {\bf k}\, \omega_K = - \frac{1}{2} dq \wedge d \bar q
\ee
where $\bar q = x_0 - {\bf i}\, x_1 - {\bf j}\, x_2 - {\bf k}\, x_3$
is the conjugate quaternion. Explicitly,
\begin{eqnarray}
\omega_I & = & dx_0 \wedge dx_1 + dx_2 \wedge dx_3 \nonumber \\
\omega_J & = & dx_0 \wedge dx_2 - dx_1 \wedge dx_3 \label{ijkonr4} \\
\omega_K & = & dx_0 \wedge dx_3 + dx_1 \wedge dx_2 \nonumber
\end{eqnarray}
Simple examples of $(B,A,A)$ branes are branes supported on $f(z,w)=0$,
where $f(z,w)$ is a holomorphic function of $z = x_0 + i x_1$ and $w = x_2 + i x_3$.
With the above definitions, it is easy to verify that these submanifolds
are complex for complex structure $I$ and Lagrangian for $\omega_J$ and $\omega_K$.
\endexample

The second key ingredient, the coisotropic brane $\Bcc$, is an $A$-brane on $Y$
with respect to $\omega_K$, but at the same time it is a $B$-brane in complex structure $I$.
In fact, in the $B$-model of $(Y,I)$ the brane $\Bcc$ corresponds to
a holomorphic line bundle that, abusing notations a little,
we also denote $\CL \to Y$, with the first Chern class
\be
c_1 (\CL) \; = \; \omega_I \,.
\label{bccfirstchern}
\ee
Therefore, when $Y$ admits a hyper-K\"ahler structure, the Hilbert space $\CH$ is simply
given by \eqref{hinbi} and can be analyzed in complex structure $I$
using the tools of algebraic geometry.\\

Our next goal is to see whether the extra structure of $Y$ being a hyper-K\"ahler
manifold and $\CB'$, $\Bcc$ being branes of type $(B,A,A)$
can help us to identify the mirror objects $\tCB'$, $\tBcc$.
As explained in \cite{KW} and as we illustrate in many examples below,
in general a brane of type $(B,A,A)$ transforms under mirror symmetry
into a brane of type $(B,B,B)$, {\it i.e.} a $B$-brane for all complex
structures on $\tilde Y$:
$$
\phantom{\oint}
\Phi_{{\rm mirror}} ~:~~~~ \quad (B,A,A) {\rm ~branes~} \; \quad \longrightarrow \quad \; (B,B,B) {\rm ~branes~}
$$
This statement depends, of course, in a crucial way on the fact that the fibers
of the SYZ fibration \eqref{syzmirror} are also of type $(B,A,A)$,
{\it i.e.} the fibration is holomorphic in complex structure $I$
and Lagrangian for $\omega_J$ and $\omega_K$.

In particular, since the fibration \eqref{syzmirror} is assumed to be holomorphic in
complex structure $I$, mirror symmetry transforms holomorphic objects into holomorphic objects
and, as a result, does not change the type of branes in complex structure~$I$.
(This, of course, is not the case in other complex structures.)
Moreover, from the vantage point of the complex structure $I$,
the SYZ duality along the fibers $\FF$ can be described as a Fourier-Mukai transform:
\be
\Phi_{{\rm FM}} ~:~~~ \CD^b (Y,I) \;\overset{\sim}{\longrightarrow}\; \CD^b (\tilde{Y}, \tilde{I}) \,,
\label{fmmap}
\ee
where, to avoid confusion, we made explicit the choice of complex structures.
This point of view can be very helpful in identifying the mirror objects \eqref{bbmirrors}
dual to our branes $\CB'$ and $\Bcc$. As long as they are $B$-branes in complex structure $I$,
their mirrors can be obtained by the following general formula
\be
\tCB = {\bf R} \tilde p_* \left( p^* \CB \otimes \CP \right) \,,
\label{fmonb}
\ee
which describes explicitly the action of the functor \eqref{fmmap} on a brane $\CB \in \CD^b (Y,I)$.
Here, $\CP$ is the relative Poincar\'e line bundle on $Z := Y \times_{\BB} \tilde Y$, and
$$
\begin{array}{ccccc}
\; & \; & Z & \; & \; \\
\; & \overset{p}{\swarrow} & \; & \overset{\tilde p}{\searrow} & \; \\
Y & \; & \; & \; & \tilde Y
\end{array}
$$
In particular, the Chern character of the mirror $(B,B,B)$ brane $\tCB$ is given by
\be
\ch (\tCB) = \tilde p_* \left( \ch (\CP) \wedge p^* (\ch (\CB)) \right) \,.
\label{chmap}
\ee
Although the viewpoint of complex structure $I$ is extremely useful
(and we shall return to it later), now we wish to proceed with a more
democratic approach where $\tBcc$ and $\tCB'$ are considered as objects of type $(B,B,B)$.
In particular, our goal is to understand what this extra structure really means
and what it can be good for.\\

The simplest example of a $(B,B,B)$ brane on $\tilde Y$ is
a {\it hyperholomorphic} bundle $E$, {\it i.e.} a holomorphic bundle
compatible with the hyper-K\"ahler structure on $\tilde Y$,
in the sense that $E$ admits a Hermitian connection $\nabla$ with
a curvature $F_{\nabla} \in \Lambda^2 (\tilde Y, {\rm End} (E))$
which is of Hodge type $(1,1)$ with respect to all complex structures.
A stable bundle $E$ is hyperholomorphic if and only if its Chern classes
$c_1$ and $c_2$ are $SU(2)$-invariant, with respect to the natural $SU(2)$ action
on the cohomology, see {\it e.g.} \cite{Verbitsky}:
\medskip
\be
\begin{array}{ccc}
\label{hyperholcriterion}
\underline{E ~\text{hyperholomorphic}}
& \quad \Leftrightarrow \quad &
\underline{c_1 (E), ~c_2 (E) \quad SU(2)\text{-invariant}} \vspace{1mm}
\end{array}
\ee
This simple criterion is our first indication that the study of $(B,B,B)$ branes
is closely related to the study of $SU(2)$ action on the cohomology of $\tilde Y$.

For instance, if $\tilde Y = \R^4$ as in our previous example, then
the left multiplication $q \mapsto u \cdot q$ by a unit quaternion $u$ ($u \bar u =1$)
is an isometry of the flat hyper-K\"ahler metric $ds^2 = dq d \bar q$ on $\tilde Y = \R^4$
and rotates the three K\"ahler forms \eqref{ijkonr4}.
This gives a rather explicit local model for $SU(2)$ action on the cohomology of $\tilde Y$.
In general, when we apply the criterion \eqref{hyperholcriterion}
to the Chern character $\ch (\tCB)$ of the brane $\tCB$ we shall often use the fact that
a differential form $\omega$ on a hyper-K\"ahler manifold $\tilde Y$
is $SU(2)$-invariant if and only if it is of Hodge type $(p,p)$
with respect to all complex structures on $\tilde Y$.

A larger class of examples of $(B,B,B)$ branes on $\tilde Y$
can be obtained by considering hyperholomorphic sheaves,
{\it i.e.} coherent sheaves compatible with a hyper-K\"ahler structure,
in the same sense as hyperholomorphic bundles are holomorphic bundles
compatible with a hyper-K\"ahler structure.

\example{$(B,B,B)$ branes on $\tilde Y = T^* {\bf S}^2$}
In quantization of $M = {\bf S}^2$ we encountered the Eguchi-Hanson metric
on a complex surface \eqref{ehcplxsurf} which, up to a hyper-K\"ahler rotation
and irrelevant technicalities, is essentially self-mirror.
Therefore, as a first approximation to $\tilde Y$ we can take
a locally asymptotically flat hyper-K\"ahler metric on $T^* {\bf S}^2$,
for which the K\"ahler forms can be written explicitly
\begin{eqnarray}
\omega_I & = & e_0 \wedge e_1 + e_2 \wedge e_3 \nonumber \\
\omega_J & = & e_0 \wedge e_2 - e_1 \wedge e_3 \label{ijkeguchih} \\
\omega_K & = & e_0 \wedge e_3 + e_1 \wedge e_2 \nonumber
\end{eqnarray}
in the orthonormal basis
$$
e_0 = f^{-1/2} dr ~,~
e_1 = \frac{r}{2} f^{1/2} (d \psi - \cos \theta d \varphi) ~,~
e_2 = \frac{r}{2} d \theta ~,~
e_3 = \frac{r}{2} \sin \theta d \varphi
$$
with $f(r) = 1 - \frac{r_0^4}{r^4}$.
This metric admits a normalisable anti-self-dual harmonic $2$-form
\be
\varpi \; = \; \frac{1}{r^4} (e_0 \wedge e_1 - e_2 \wedge e_3) \,,
\label{ehasd2form}
\ee
which, according to \eqref{hyperholcriterion}, can represent the first Chern class
of a $(B,B,B)$ brane $\tCB$. Indeed, the $2$-form \eqref{ehasd2form}
is of type $(1,1)$ with respect to all complex structures on $\tilde Y \cong T^* {\bf S}^2$
and is orthogonal to all three K\"ahler forms \eqref{ijkeguchih}.
\endexample

This example is the simplest case of the following infinite family
of hyper-K\"ahler metrics on $T^* \cp^n$ discovered by E.~Calabi \cite{Calabi}
(who also introduced the term ``hyper-K\"ahler'').

\example{$(B,B,B)$ branes on $\tilde Y = T^* \cp^n$}
In an orthonormal basis of $1$-forms, the Calabi metric has the standard form
$$
ds^2 \; = \; \sum_{i=1}^n \sum_{a=0}^3 e^{(i)}_a \otimes e^{(i)}_a
$$
with the K\"ahler forms, {\it cf.} \eqref{ijkeguchih},
\begin{eqnarray}
\omega_I & = & \sum_{i=1}^n \left( e^{(i)}_0 \wedge e^{(i)}_1 + e^{(i)}_2 \wedge e^{(i)}_3 \right) \nonumber \\
\omega_J & = & \sum_{i=1}^n \left( e^{(i)}_0 \wedge e^{(i)}_2 - e^{(i)}_1 \wedge e^{(i)}_3 \right) \label{ijkcalabi} \\
\omega_K & = & \sum_{i=1}^n \left( e^{(i)}_0 \wedge e^{(i)}_3 + e^{(i)}_1 \wedge e^{(i)}_2 \right) \nonumber
\end{eqnarray}
These K\"ahler forms are rotated by the $SU(2)$ symmetry,
under which the basis $1$-forms transform as doublets:
$$
\begin{pmatrix}
e^{(i)}_0 + i e^{(i)}_1 \\ e^{(i)}_2 - i e^{(i)}_3
\end{pmatrix}
\qquad {\rm and} \qquad
\begin{pmatrix}
e^{(i)}_2 + i e^{(i)}_3 \\ - e^{(i)}_0 + i e^{(i)}_1
\end{pmatrix}
\,.
$$
{}From these one can construct singlets, {\it i.e.} $SU(2)$-invariant forms on $\tilde Y$,
which include the following harmonic 2-form \cite{CGLP}:
\be
\varpi \; = \;
\frac{1}{r^4} \left( e^{(1)}_0 \wedge e^{(1)}_1 - e^{(1)}_2 \wedge e^{(1)}_3 \right)
+ \frac{1}{r^2} \sum_{i=2}^n \left( e^{(i)}_0 \wedge e^{(i)}_1 - e^{(i)}_2 \wedge e^{(i)}_3 \right) \,.
\label{calabi2form}
\ee
This harmonic 2-form is not normalisable (except for $n=1$, when it reduces to \eqref{ehasd2form}),
but it is regular and square-integrable at $r=r_0$.
\endexample

The special case ($n=3$) of this last example shows up in the $B$-model approach
to quantization of $M = \M_{{\rm flat}} (G,C)$, with $G=SO(3)$ and $C$ of genus $g=2$;
see comments below \eqref{dimhgenus2} and section \ref{sec:cstheory}.
In particular, the $SU(2)$-invariant 2-form \eqref{calabi2form} turns out to be
essentially the first Chern class of the $(B,B,B)$ brane~$\tBcc$.\\

Another way to construct a $(B,B,B)$ brane is to take
an ideal sheaf of a trianalytic subvariety of $\tilde Y$.
Trianalytic subvarieties have an action of quaternion algebra in the tangent bundle.
In particular, the real dimension of such subvarieties is divisible by 4.
By analogy with hyperholomorphic bundles (sheaves) they can be characterized
by the following criterion, similar to \eqref{hyperholcriterion}:
if $S \subset \tilde Y$ is a closed analytic subvariety of $(\tilde Y, \tilde{I})$
and $[S] \in H^{4i} (\tilde Y)$ is $SU(2)$-invariant, then $S$ is trianalytic.
Trianalytic subvarieties are quite rare;
for example, a Hilbert scheme of a generic $K3$ surface provides a good example
of a compact hyper-K\"ahler manifold, but it has no trianalytic subvarieties.\\

In order to keep our discussion less abstract, in the rest of this section we work out
in detail two simple, yet non-trivial examples based on $\tilde Y$ of (real) dimension~$4$.
Clearly, in these examples $\ch (\tCB)$ has components only in degree $0$, $2$, and $4$,
so that \eqref{hyperholcriterion} provides a non-trivial constraint only
on a degree-$2$ component, {\it i.e.} on the first Chern class of $\tCB$.
On the other hand, if $\omega_I$, $\omega_J$, $\omega_K$, $\varpi_1, \ldots, \varpi_k$ is an orthonormal
basis in $H^2 (\tilde Y)$, then the vectors $\varpi_1, \ldots, \varpi_k$ are $SU(2)$-invariant,
and in the natural $SU(2)$-invariant decomposition
\be
H^2 (\tilde Y) = H^2_{{\rm inv}} (\tilde Y) \oplus H^2_+ (\tilde Y)
\label{hsdhasd}
\ee
we have $\dim H^2_+ (\tilde Y)=3$ and $H^2_{{\rm inv}} (\tilde Y) \cong H^2_- (\tilde Y)$,
where $H^2 (\tilde Y) = H^2_+ (\tilde Y) \oplus H^2_- (\tilde Y)$
is the standard decomposition of $H^2 (\tilde Y)$ according to
the eigenvalues of the Hodge $*$ operator.


\subsection{Toy model}
\label{sec:torus}

In this section, we apply the general formalism described above to a simple model,
where the SYZ fibration \eqref{syzmirror} is actually trivial. Specifically, we take
\be
Y \; = \; \BB \times \FF
\ee
where $\BB = \R^2$ and $\FF = T^2$.
This model can be regarded as a quantization of $M = T^2$, {\it cf.} \eqref{yccstar}.
Indeed, if we choose $\CB' = \CB_{\FF}$ to be a Lagrangian brane supported on $M = \FF$
and $\Bcc$ to be a coisotropic brane with the appropriate Chan-Paton bundle~$\CL$,
we obtain precisely the setup of section \ref{sec:amodel}.
Note, this Lagrangian brane $\CB'$ is the one we also used in \eqref{syzfp}
to prove the general formula \eqref{tbccrank}. In particular, in \eqref{dimhtorus}
we already calculated the dimension of the corresponding Hilbert space~$\CH$.

As a warm-up to more interesting models, we wish to show explicitly in this example
that the branes $\CB'$ and $\Bcc$ are compatible with the hyper-K\"ahler structure on $Y$
and to use this information to find the mirror $(B,B,B)$ branes $\tCB'$ and $\tBcc$.
In order to do this, however, we first need to introduce the complex structures $I$, $J$, $K$,
and the corresponding K\"ahler forms on $Y$.
These are essentially written in \eqref{ijkonr4}.
Let $b_1$, $b_2$ be a basis of 1-forms on the base $\BB$,
and $f_1$, $f_2$ (resp. $\tilde f_1$, $\tilde f_2$) be a basis of 1-forms on the fiber $\FF$
(resp. the dual fiber $\widetilde{\FF}$).
Then, the K\"ahler forms on $Y$ are
\begin{eqnarray}
\omega_I & = & \frac{1}{\hbar} \left( b_1 \wedge b_2 + f_1 \wedge f_2 \right) \nonumber \\
\omega_J & = & \frac{1}{\hbar} \left( b_1 \wedge f_1 - b_2 \wedge f_2 \right) \label{ijktoy} \\
\omega_K & = & \frac{1}{\hbar} \left( b_1 \wedge f_2 + b_2 \wedge f_1 \right) \nonumber
\end{eqnarray}
where, compared to \eqref{ijkonr4}, we introduced the parameter $\hbar$ relevant for quantization.

Dualizing the fiber $\FF$, we obtain the mirror manifold
\be
\tilde Y \; = \; \BB \times \widetilde{\FF} \,,
\ee
which, of course, is also a trivial SYZ fibration,
with the fiber $\widetilde{\FF} = H^1 (\FF, U(1)) \cong T^2$.
Moreover, since the $U(1)^2$ isometry of $Y$ (that acts in
a natural way by translations along the SYZ fibers)
is tri-holomorphic, the duality certainly does not spoil
the hyper-K\"ahler structure.
Hence, the resulting mirror manifold $\tilde Y$ is also
hyper-K\"ahler, and the corresponding K\"ahler forms
\begin{eqnarray}
\tilde \omega_I & = & \frac{1}{\hbar} b_1 \wedge b_2 + \hbar \tilde f_1 \wedge \tilde f_2 \nonumber \\
\tilde \omega_J & = & ~b_1 \wedge \tilde f_1 - b_2 \wedge \tilde f_2 \label{ijktoymirror} \\
\tilde \omega_K & = & ~b_1 \wedge \tilde f_2 + b_2 \wedge \tilde f_1 \phantom{\oint}~ \nonumber
\end{eqnarray}
can be obtained from \eqref{ijktoy} simply by replacing $f_i \to \hbar \tilde f_i$.
Note, in particular, that in the K\"ahler metric corresponding to
the complex structure $I$, we have ${\rm Vol} (\FF) \sim \hbar^{-n}$,
where $n = \dim_{\C} \FF$, and ${\rm Vol} (\widetilde{\FF}) \sim \hbar^n$.
This property holds true for more general mirror pairs, $Y$ and $\tilde Y$.\\

Now, we wish to identify the mirror $(B,B,B)$ branes $\tCB'$ and, most importantly,~$\tBcc$.
The brane $\tCB'$ is easy to identify and, in fact, we already took care of it
in our earlier discussion: as summarized in \eqref{syzfp},
mirror symmetry maps a Lagrangian brane $\CB_{\FF}$
to a skyscraper sheaf $\CB_p = \CO_p \in \CD^b (\tilde Y)$.
Therefore, if we choose $\CB' = \CB_{\FF}$, as in our approach to
quantization of $M = T^2$, then the mirror $B$-brane is $\tCB' = \CB_p$.
It has the Chern character
\be
\ch (\tCB') \; = \; - b_1 \wedge b_2 \wedge \tilde f_1 \wedge \tilde f_2 \,,
\label{chbzerotoy}
\ee
which is consistent\footnote{We leave this as an exercise.} with \eqref{chmap}
and is manifestly invariant under the $SU(2)$ action on the cohomology of $\tilde Y$.
(Clearly, the degree-$0$ form and the volume form are $SU(2)$-invariant
on any hyper-K\"ahler manifold $\tilde Y$.)

Identifying the mirror of the coisotropic $(B,A,A)$ brane $\Bcc$ is more interesting.
On general grounds, we know that it should be an object of type $(B,B,B)$,
{\it i.e.} holomorphic in all complex structures on $\tilde Y$,
and, according to \eqref{tbccrank},
should have $\rank (\tBcc) = \int_{\FF} \omega_I = \frac{1}{\hbar}$.
Therefore, we expect
\be
\ch (\tBcc) = \frac{1}{\hbar} + \ldots \,,
\label{bccguesstoy}
\ee
where the rest of the terms (denoted by ellipsis)
should be invariant under the $SU(2)$ action on the cohomology of~$\tilde Y$.
Besides the $0$-form and the volume form (which are always $SU(2)$-invariant),
such terms may contain any linear combination of the anti-self-dual 2-forms on $\tilde Y$:
\be
\frac{1}{\hbar} b_1 \wedge b_2 - \hbar \tilde f_1 \wedge \tilde f_2 \,, \qquad
b_1 \wedge \tilde f_1 + b_2 \wedge \tilde f_2 \,, \qquad
b_1 \wedge \tilde f_2 - b_2 \wedge \tilde f_1 \,,
\label{asdformstoy}
\ee
which, according to \eqref{hsdhasd}, are precisely the generators
of the $SU(2)$-invariant part of the cohomology $H^2 (\tilde Y)$.
(It is easy to verify that all of the forms in \eqref{asdformstoy}
are orthogonal to the K\"ahler forms \eqref{ijktoymirror}.)
Of course, this structure alone does not uniquely determine $\ch (\tBcc)$,
but it is amusing to see how close we managed to get to the correct answer.

In order to compute $\ch (\tBcc)$ more systematically, we can treat
the coisotropic brane $\Bcc$ as a $B$-brane in complex structure $I$,
where it corresponds to a holomorphic line bundle $\CL$
with the first Chern class \eqref{bccfirstchern}.
Then, substituting $\ch (\Bcc) = \exp (\omega_I)$
into the general formula \eqref{chmap},
\be
\ch (\tBcc) \; = \; \int_{\FF} \ch(\CP) \wedge  p^* (\ch (\Bcc)) \,,
\label{bccchmap}
\ee
where $\CP$ is a complex line bundle on $Z = \BB \times \FF \times \widetilde{\FF}$
defined by its first Chern class,
\be
c_1 (\CP) \; = \; \sum_{i=1}^2 f_i \wedge \tilde f_i \,,
\ee
we obtain the Chern character of the mirror $(B,B,B)$ brane $\tBcc$:
\be
\ch (\tBcc) \; = \; \frac{1}{\hbar} + \frac{1}{\hbar^2} b_1 \wedge b_2
- \tilde f_1 \wedge \tilde f_2 - \frac{1}{\hbar} b_1 \wedge b_2 \wedge \tilde f_1 \wedge \tilde f_2 \,.
\label{chbccmirrortoy}
\ee
Note, the degree-$0$ term in this expression is precisely what
we found in \eqref{bccguesstoy} and the first Chern class is (a multiple of)
one of the $SU(2)$-invariant 2-forms \eqref{asdformstoy}.
Therefore, we conclude that, in the present example,
$\tBcc$ is a hyperholomorphic bundle on $\tilde Y$
with the Chern character \eqref{chbccmirrortoy} which,
in accordance with the criterion \eqref{hyperholcriterion},
is invariant under the $SU(2)$ action on the cohomology of~$\tilde Y$.

Now, if we wish to return to the original quantization problem,
there is a simple way to obtain the Hilbert space $\CH$ associated
with the quantization of $M=T^2$ directly in the $B$-model of $\tilde Y$.
In the present case, only $\Ext^0_{\tilde Y} (\tBcc,\tCB')$
contributes to \eqref{mirrorhbbcc} and its dimension can be
found with the help of the Grothendieck-Riemann-Roch theorem \eqref{GRRtheorem}:
\be
\dim \CH = \dim \Ext^0_{\tilde Y} (\tBcc,\tCB')
= \int_{\widetilde{Y}} \ch (\tBcc)^* \wedge \ch (\tCB') = \frac{1}{\hbar} \,,
\label{torusbmodel}
\ee
where we used \eqref{chbzerotoy} and \eqref{chbccmirrortoy}.
In fact, as we already pointed out earlier, the dimension of $\CH$
in this example was already computed in \eqref{dimhtorus}
when we discussed the rank of $\tBcc$.

In our next example, we consider a mirror pair, $Y$ and $\tilde Y$,
also of (real) dimension $4$, but with a non-trivial SYZ fibration \eqref{syzmirror}.


\subsection{$(B,A,A)$ and $(B,B,B)$ branes on $K3$}

The first non-trivial example of a (compact) hyper-K\"ahler manifold
is a $K3$ surface. This example is special in a number of ways.
In particular, it is the first (and so far the only) example
of a compact Calabi-Yau manifold $Y$ of $\dim_{\C} Y > 1$,
for which the homological mirror symmetry \eqref{mirrormap} is actually a theorem.
In this case, the mirror manifold $\tilde Y$ is also a $K3$ surface,
so that some of the discussion below should apply equally well to both
$Y$ and $\tilde Y$.

We remind that, topologically, a $K3$ surface is a compact
simply-connected 4-manifold,
with non-trivial Betti numbers $b_0 = 1$, $b_2 = 22$, and $b_4 = 1$.
Its cohomology group $H^2(Y, \Z)$ is an even unimodular lattice of signature $(3,19)$:
\be
\Gamma^{19,3} \; = \; (- E_8) \oplus (- E_8) \oplus U \oplus U \oplus U \,,
\label{k3lattice}
\ee
where $U$ denotes the two-dimensional even unimodular lattice $U \cong II^{1,1}$
with the intersection form
\be
\begin{pmatrix}
0 & 1 \\ 1 & 0
\end{pmatrix}
\label{hyperblattice}
\ee
and $E_8$ is the root lattice of the Lie algebra of the same name.
To make a contact with the SYZ approach to mirror symmetry \eqref{syzmirror},
we choose $Y$ (and $\tilde Y$) to be an elliptically fibered $K3$ surface
with a section, {\it i.e.} there is a map
\be
\pi : Y \to \BB \,,
\label{k3fibration}
\ee
whose general fibers are smooth elliptic curves $\FF \cong T^2$, and $\BB \cong \cp^1$.

In order to find the precise map between $(B,A,A)$ branes on $Y$
and $(B,B,B)$ branes on $\tilde Y$, it is convenient to work in complex structure $I$
(resp. $\tilde I$) where the mirror map \eqref{mirrormap} is simply the Fourier-Mukai transform \eqref{fmmap}.
Then, on both sides of mirror symmetry we deal with $B$-branes, which
can be described in terms of coherent sheaves on $Y$ and $\tilde Y$, respectively.
Given a sheaf $\CB$ on $Y$ (similarly, on $\tilde Y$) we write its Chern class as a triple
\be
(\rank (\CB), c_1 (\CB), c_2 (\CB)) \; \in \; \Z \times {\rm NS} (Y) \times \Z
\label{cherntriple}
\ee
where ${\rm NS} (Y)$ is the N\'eron-Severi lattice of $Y$,
{\it i.e.} a sublattice of $H^2(Y, \Z)$ spanned by
the cohomology classes dual to algebraic cycles of $Y$.
As a group, ${\rm NS} (Y)$ is isomorphic to the Picard group of $Y$,
that is the group of algebraic equivalence classes of holomorphic line bundles over $Y$.
The rank of the N\'eron-Severi lattice, denoted by $\rho_Y$ varies between 0 and 20,
and by the Hodge index theorem, the signature of ${\rm NS} (Y)$ is $(1, \rho_Y - 1)$.
A generic $K3$ surface has rank $\rho_Y = 0$, but for elliptic $K3$ surfaces
with section\footnote{A generic elliptically fibered $K3$ surface with a section
has $\rho_Y = 2$. The Picard number can jump further to $\rho_Y > 2$
on special subvarieties in the moduli space of complex structures on $Y$,
either if there are rational curves in the singular fibers
of the fibration \eqref{k3fibration},
or if the rank of the Mordell-Weil group jumps.}
the Picard number $\rho_Y$ is at least $2$.

Indeed, there are two special classes $\FF, \BB \in {\rm NS} (Y)$
associated to the elliptic fiber and the section.
These classes are independent and span a rank-2 sublattice in \eqref{k3lattice}
with the intersection form (in the basis $\{ \FF, \BB \}$):
\be
\begin{pmatrix}
0 & 1 \\ 1 & -2
\end{pmatrix}
\label{k3fbintersection}
\ee
It is easy to see that the null vectors $e_1 = \FF$ and $e_2 = \FF + \BB$
generate the two-dimensional hyperbolic lattice $U = \langle e_1, e_2 \rangle$
with the intersection form \eqref{hyperblattice}.
Mirror symmetry identifies this lattice with another copy
of the two-dimensional hyperbolic lattice, $U \cong H^0(\tilde Y, \Z) \oplus H^4(\tilde Y, \Z)$.
Indeed, as described in \eqref{cherntriple} the Chern classes of branes
on $Y$ and $\tilde Y$ take values in the lattice $\Z \times {\rm NS} \times \Z$,
which, for generic $K3$ surfaces in the class that we consider, is a lattice of rank $4$.
Specifically, this lattice is isomorphic to $U \oplus U$,
and mirror symmetry acts by exchanging the two copies of $U$.

Our next goal is to see more explicitly how mirror symmetry acts
on particular branes. Of course, we are especially interested in
the coisotropic brane $\Bcc$, which in the $B$-model of $(Y,I)$
corresponds to a holomorphic line bundle $\CL \to Y$
with the first Chern class \eqref{bccfirstchern}.
Clearly, this line bundle (and the brane $\Bcc$) can only exist
if $c_1 (\CL) = \omega_I$ is an element in ${\rm NS} (Y)$, in other words,
only if\footnote{Here, we assume a generic situation with $\rho_Y = 2$.}
\be
\omega_I = k \BB + k' \FF
\ee
for a pair of integer numbers $k >0$ and $k' \gg 0$ that,
by analogy with \eqref{khinteger}, we shall call the ``level.''
Assuming this is the case, and applying the Fourier-Mukai transform \eqref{fmonb}
to $\CL$, we obtain the dual bundle (sheaf) on $\tilde Y$
with the Chern character, {\it cf.}~\eqref{chmap},
\begin{eqnarray}
\ch_0 (\tBcc) & = & k \nonumber \\
\ch_1 (\tBcc) & = & (k' - k) k \tilde \FF - \tilde \BB \label{chbccmirrork3} \\
\ch_2 (\tBcc) & = & - k' \nonumber
\end{eqnarray}
In general, this answer describes a higher rank object $\tBcc$ on $\tilde Y$
and has a structure similar to \eqref{chbccmirrortoy}.
Compared to \eqref{chbccmirrortoy}, however, it has some extra ``corrections''
due to non-trivial geometry of the fibration \eqref{k3fibration} in our present example.

Now, let us take a closer look at the properties of the $(B,B,B)$ brane $\tBcc$.
First, we recall that
the moduli space of coherent semi-stable shaves on $\tilde Y = K3$
with fixed Chern classes is a smooth and compact manifold of dimension \cite{Mukai}:
\be
\dim_{\R} \CM (\CB) = 2 v^2 + 4
\label{mukaidim}
\ee
where $v = v (\CB)$ is the charge vector of a brane $\CB$
($=$ the Mukai vector of the corresponding coherent sheaf):
\be
\begin{array}{ccccccc}
v (\CB) := \ch (\CB) \sqrt{{\rm Td} (\tilde Y)} & = & r & + & c_1 & + & \ell \\
& \in & H^0 & \oplus & H^2 & \oplus & H^4 \\
\; & \; & D4 & \; & D2 & \; & D0
\end{array}
\label{mukaicharge}
\ee
Explicitly, in \eqref{mukaidim} the inner product
of the Mukai vector \eqref{mukaicharge} is given by
\be
v^2 = c_1^2 - 2r \ell \,,
\ee
where $c_1^2$ is the inner product on $H^2 (\tilde Y,\Z) \cong \Gamma^{19,3}$.
Applying this to the brane $\tBcc$ with the Chern character \eqref{chbccmirrork3}
and using \eqref{k3fbintersection}, we find $\dim \CM (\tBcc) = 0$.
We expect this to be a general feature of the mirror
of the canonical coisotropic brane.

\begin{conjecture}
The brane $\tBcc$ is always rigid.
\end{conjecture}

Returning to the original quantization problem, now we are ready
to quantize any symplectic 2-manifold $M \subset Y$, on which
$\omega = \omega_I \vert_M$ is non-degenerate.
As usual, we take $\CB'$ to be a Lagrangian brane supported on $M$ and,
to make use of the hyper-K\"ahler structure on $Y$, we choose $M$ to be
analytic in complex structure $I$ and Lagrangian for $\omega_J$ and $\omega_K$.
Then, $\CB'$ is a brane of type $(B,A,A)$ supported\footnote{As in the earlier
discussion, we assume that $Y$ is a generic elliptically fibered $K3$ surface
with the N\'eron-Severi lattice ${\rm NS} (Y)$ of rank $\rho_Y =2$
generated by the classes $\FF$ and $\BB$.} on a holomorphic curve
in the homology class $M = n_{\FF} \FF + n_{\BB} \BB$, whose genus follows from the adjunction formula
\be
2g (M) - 2 \; = \; M \cdot M \,,
\ee
and the intersection pairing \eqref{k3fbintersection}.
Applying the Fourier-Mukai transform \eqref{fmonb},
it is easy to see that the mirror $(B,B,B)$ brane $\tCB'$
is described by a hyperholomorphic sheaf on $\tilde Y$
with the Chern character
\be
\ch (\tCB') = (n_{\BB}, 0, -n_{\FF}) \,.
\label{chbzerok3}
\ee
This answer is manifestly invariant under the $SU(2)$ action on
the cohomology of~$\tilde Y$, in perfect agreement with the general  criterion \eqref{hyperholcriterion}.
To make contact with the quantization of a 2-torus $T^2$
considered in \eqref{yccstar} and then in more detail in section \ref{sec:torus},
we can take $M = \FF$ to be a copy of the fiber.
Then, the $B$-model approach \eqref{mirrorhbbcc} leads to a Hilbert space $\CH$
of dimension  $\dim \CH = k = \frac{1}{\hbar}$,
in agreement with earlier results \eqref{hdimvolm}, \eqref{dimhtorus}, and \eqref{torusbmodel}
that we already rederived several times in this paper from various angles.


\subsection{What became of $\hbar$}

In the original quantization problem, $\hbar$ determines the norm of
the symplectic form $\omega$ on the symplectic manifold $M$.
After embedding the quantization problem in the $A$-model of $Y$,
the closed 2-form $\omega$ and, therefore, the parameter $\hbar$
acquire a new interpretation.
Namely, $\omega$ becomes (the restriction to $M \subset Y$ of)
the curvature 2-form $F$ of a unitary line bundle $\CL \to Y$,
the Chan-Paton bundle of~$\Bcc$.
Mirror symmetry maps the coisotropic brane $\Bcc$ to a $B$-brane $\tBcc$,
and $\hbar$ determines the topology of its Chan-Paton bundle,
{\it cf.} \eqref{chbccmirrortoy} and \eqref{chbccmirrork3}.

Since the definition of the branes $\Bcc$ and $\tBcc$ is intimately tied
with the geometry of $Y$ and $\tilde Y$,
the parameter $\hbar$ also admits a purely geometric interpretation,
either as a symplectic structure of $Y$ or, via mirror symmetry,
as a complex structure of $\tilde Y$.
In our toy model of section \ref{sec:torus} this is easy to see from
the explicit formulas \eqref{ijktoy} and \eqref{ijktoymirror}.

As illustrated in \eqref{quantizationmap}, after quantization the parameter
$\hbar$ enters the definition of various quantum objects, such as $\CH$ and $\CA_{\hbar}$.
In particular, when the phase space $M$ is compact,
the Hilbert space $\CH$ is finite-dimensional,
and $\hbar$ determines the dimension of $\CH$,
as in the volume integrals \eqref{dimhsphere}, \eqref{hdimvolm} or \eqref{dimhtorus}.

What happens if $M$ is non-compact?
For example, in our toy model of section \ref{sec:torus}
we could just as well take $M$ to be a copy of $\BB = \R^2$
(embedded in $Y = \BB \times \FF$ in an obvious way).
Then, the Hilbert space $\CH$ is infinite-dimensional,
and the closest to $\dim \CH$ is the trace,
\be
\Tr_{\CH} e^{- \beta H} \,,
\label{htrace}
\ee
that one can define by introducing a Hamiltonian $H$
and a parameter $\beta$.
Classically, $H$ is simply a function on $M$.
According to the general principle \eqref{quantizationmap},
after quantization it becomes an operator on $\CH$,
and the partition function \eqref{htrace} encodes the spectrum of $H$.
Note, when $M$ is compact,
\eqref{htrace} gives the dimension of $\CH$ if we set $\beta$ or $H$ to zero.

\example{Quantization of $M = \R^2$}
This problem can be realized as a special case of our toy model
in section \ref{sec:torus}, if we take $M = \BB$.
In the $A$-model approach, the symplectic form $\omega$
is the restriction to $M \subset Y$ of the K\"ahler form $F = \omega_I$,
\be
\omega = \omega_I \vert_M = \frac{1}{\hbar} dx_0 \wedge dx_1 \,,
\label{rtwosympl}
\ee
where $x_0$ and $x_1$ are linear coordinates on $\BB$,
{\it cf.} \eqref{ijkonr4} and \eqref{ijktoy}.
Introducing the Hamiltonian $H = \frac{1}{2} (x_0^2 + x_1^2)$,
we obtain a classical example of a quantum system, namely the quantum harmonic oscillator.
The eigenvalues of this Hamiltonian are
\be
H_i = \left( i + \frac{1}{2} \right) \hbar \,,
\qquad i = 0, 1, 2, \ldots
\ee
so that one can easily perform the sum in \eqref{htrace} to obtain the partition function
\be
\Tr_{\CH} e^{- \beta H} = \frac{e^{-\beta \hbar/2}}{1 - e^{-\beta \hbar}}
= \frac{1}{2 \sinh (\beta \hbar/2)} \,.
\label{harmosc}
\ee
Note, that $\hbar$ appears only in a combination with $\beta$.
\endexample

Just as in the finite-dimensional case,
the trace \eqref{htrace} is closely related to a volume integral
of the form
\be
\int_M e^{F - \beta H} \; = \; \int_M \frac{F^n}{n!} e^{-\beta H}
\label{eqvolume}
\ee
that, in favorable situations, one can also interpret as
the ``equivariant volume'' of~$M$.
Indeed, if $M$ (resp. $Y$) admits a circle action,
which is Hamiltonian\footnote{The action of $G$ on $M$ is called Hamiltonian
with moment map $\mu : M \to \frak g^*$ if $d \langle \beta, \mu \rangle = - \iota (V \beta) \cdot \Omega$
for every $\beta \in \frak g$,
where $\langle \; , \; \rangle$ denotes the pairing between $\frak g$ and $\frak g^*$.
This implies, among other things, that the zeroes of the vector field $V \beta$
are precisely the critical points of $\langle \beta, \mu \rangle$.} with moment map $H : M \to \R$,
then the combination $F - \beta H$ that appears in the exponent
of \eqref{eqvolume} can be interpreted as the equivariant
symplectic form on $M$ (resp. $Y$), provided we identify $\beta$
with the generator of the base ring
\be
H_{{\bf S}^1}^* ({\rm pt}) = H^* (\cp^{\infty}) = \C [\beta] \,.
\ee
Indeed, since the ${\bf S}^1$ action is Hamiltonian and $F$ is closed,
we have $(d - \iota (V \beta) )(F - \beta H) = 0$,
so that $F - \beta H$ is a closed equivariant form.
In fact, it is the equivariant first Chern class
of a complex line bundle $\CL$ compatible with the circle action.
Therefore, the integrand of \eqref{eqvolume} is simply the equivariant Chern character of~$\CL$.
Other characteristic classes also can be extended to ${\bf S}^1$-equivariant forms.
For example, the Todd class --- which often accompanies Chern characters in our
integration formulas --- can be combined with $\ch (\CL, \beta)$ to produce
an equivariant version of the integral in the Riemann-Roch formula \eqref{GRRtheorem},
\be
\int_M \ch (\CL,\beta) \wedge {\rm Td} (M, \beta) \,,
\label{eqRR}
\ee
that, in the equivariant setting, computes the ${\bf S}^1$-equivariant
index of the Spin$^c$ Dirac operator $\slash \!\!\! \partial_{\CL}$,
twisted by $\CL$ \cite{BGV}.

The equivariant integrals \eqref{eqvolume} and \eqref{eqRR}
localize on the fixed points of the circle group action
({\it i.e.} zeroes of the corresponding vector field $V$).
Thus, the Duistermaat-Heckman formula asserts that
the equivariant symplectic volume \eqref{eqvolume}
can be written as a sum of local contributions of the fixed points
(which, for simplicity, we assume to be isolated):
\be
\int_M \frac{F^n}{n!} e^{-\beta H} \; = \;
\sum_{p \, \in \, {\rm zeroes \, of} \, V} \; \frac{e^{-\beta H(p)}}{\beta^n e (p)} \,,
\label{DHformula}
\ee
where $e (p) = w_1 \ldots w_n$ is the product of the weights of the ${\bf S}^1$ action on $T_p M$.
Similarly, by the Atiyah-Segal-Singer equivariant index theorem \cite{ASS},
the ${\bf S}^1$-equivariant index of the Spin$^c$ Dirac operator $\slash \!\!\! \partial_{\CL}$
can be expressed as a localization of the integral \eqref{eqRR},
\be
{\rm index}_{{\bf S}^1} (\slash \!\!\! \partial_{\CL})
\; = \; \sum_s \int_{F_s} \frac{\ch (\CL,\beta) \, {\rm Td} (F_s, \beta)}{\prod (1-e^{-x_i-\beta w_i})} \,,
\label{ASSformula}
\ee
where the sum runs over connected components of the fixed point set of ${\bf S}^1$,
and $x_i$, $i = 1, \ldots, \frac{1}{2} {\rm codim} F_s$, are the formal Chern roots
of the normal bundle of $F_s$.

\example{Quantization of $M = \R^2$}
Continuing with our previous example, we wish to study the space~$\CH$
of open strings between the coisotropic brane $\Bcc$ on $Y = \BB \times \FF$
and the Lagrangian $A$-brane $\CB'$ supported on $M = \BB \times \{ {\rm pt} \}$.
Unlike the space discussed in section \ref{sec:torus}, $\CH$ is infinite-dimensional now,
so we shall analyze it using the equivariant technique and compare the result with \eqref{harmosc}.
To do this, we consider the standard action of the circle group ${\bf S}^1$ on $M = \R^2$,
generated by the vector field $V = x_1 \partial_{x_0} - x_0 \partial_{x_1}$.
Clearly, the origin $(x_0,x_1) = (0,0)$ is an isolated fixed point of this ${\bf S}^1$ action,
and $H = \frac{1}{2} (x_0^2 + x_1^2)$ is the Hamiltonian function for the vector field
$V$ and the symplectic form \eqref{rtwosympl}.
Now, the equivariant volume \eqref{eqvolume} can be easily evaluated:
\be
\int_M e^{F - \beta H}
= \frac{1}{2\pi \hbar} \int_{\R^2} e^{-\frac{1}{2} \beta (x_0^2 + x_1^2)} \; dx_0 dx_1
= \frac{1}{\beta \hbar} \,,
\label{eqvolrtwo}
\ee
and the result agrees, of course, with what one could find by using the
the Duistermaat-Heckman formula \eqref{DHformula}.
Notice, the expression \eqref{eqvolrtwo} describes
the $\beta \hbar \to 0$ limit of the partition function \eqref{harmosc}
and the equivariant index \eqref{ASSformula}, which in the present
case takes the form
\be
\chi_{{\bf S}^1} (\Bcc,\CB')
\; = \; \frac{1}{1-e^{-\beta \hbar}} \,.
\label{eqindexrtwo}
\ee
Both \eqref{eqvolrtwo} and \eqref{eqindexrtwo} depend only on the combination $\beta \hbar$.
\endexample

In general, the equivariant symplectic volume \eqref{eqvolume} describes
the asymptotic behavior (as $\hbar \to 0$) of the trace \eqref{htrace},
which can be viewed as a regularized version of $\dim \CH$.
This is similar to the role volume of $M$ plays in \eqref{hdimvolm}.
In order to get a better approximation to \eqref{htrace},
one can consider the equivariant index \eqref{ASSformula}
of a Dirac operator\footnote{Sometimes,
in the literature $M$ is ``quantized'' by attaching to it
the virtual vector space
$Q(M) := {\rm ker} \slash \!\!\! \partial_{\CL} - {\rm coker} \slash \!\!\! \partial_{\CL}$,
whose equivariant character is \eqref{ASSformula}.
This space, however, should not be confused with $\CH$.}
which, roughly speaking, is a ``square root''
of the second order operator whose spectrum is described by \eqref{htrace},
{\it cf.} \eqref{harmosc} and \eqref{eqindexrtwo} in our toy model example.
The upshot is that the equivariant cohomology of $Y$ (resp. $\tilde Y$)
fits in well with the $A$-model (resp. $B$-model) approach to quantization
and can be a very useful tool, especially when $\CH$ is infinite-dimensional.


\section{Quantization of Chern-Simons theory}
\label{sec:cstheory}

\hfill{\vbox{\hbox{\it I think I can safely say that nobody understands}
\hbox{\it quantum mechanics.}}}

\hfill{\vbox{\hbox{Richard Feynman}}}

\medskip


The Hilbert space of Chern-Simons theory on a Riemann surface $C$
is obtained by quantizing the moduli space of flat connections \cite{Witten}.
Therefore, we take
\be
M \; = \; \M_{{\rm flat}} (G,C) \,.
\ee
As explained in \eqref{yflatcplx},
this space has a natural complexification obtained by replacing the compact
Lie group $G$ by its complexification $G_{\C}$.

The resulting space, $Y = \M_{{\rm flat}} (G_{\C},C)$ is, in fact,
a hyper-K\"ahler manifold and can be realized as the moduli space
of Higgs bundles on $C$ with structure group $G$
(also known as the Hitchin moduli space) \cite{Hitchin}:
\be
Y \; \cong \; \MH (G,C) \,.
\label{ymhhiggs}
\ee
In order to approach the quantization problem of $M$ via mirror symmetry,
we first need to find a mirror $\widetilde{Y}$ of $Y$.
A nice fact that will be useful to us\footnote{Throughout
the paper we tacitly suppress one important detail: in general,
the moduli space $M$ (resp. its complexification $Y$) has several connected
components, which correspond to gauge bundles $E \to C$ of different topology.
Specifically, these connected components
are labeled by an element of $\pi_1 (G)$ that was denoted by $\xi$ in \cite{Ramified}.
Mirror symmetry maps $\xi$ to an element of $\CZ (\LG) \cong \pi_1 (G)$
which, similarly, labels flat $B$-fields on $\tilde Y = \MH (\LG,C)$.
For example, for $G=SU(2)$ we have $\LG = SO(3)$ and there are two choices,
classified by $\CZ (G) \cong \pi_1 (\LG) \cong \Z_2$.
If $M$ is identified, by a theorem of Narasimhan and Seshadri,
with the moduli space of (semi-)stable rank-2 bundles over $C$,
then the two choices of $\xi \in \Z_2$ correspond to bundles of even (resp. odd) degree.
In this paper we tacitly make the choice $\xi = 0$,
which corresponds to stable $G_{\C}$ bundles of even degree.}
is that $\widetilde{Y}$
is also a Hitchin moduli space, $\MH (\LG,C)$, but for the Langlands dual group $\LG$.
In fact, the mirror manifolds $\MH (G,C)$ and $\MH (\LG,C)$
fiber over the same vector space $\BB$ (under the Hitchin maps),
and the generic fibers are dual tori \cite{BJSV,HT}
(so these two fibrations give us an example of SYZ T-duality \eqref{syzmirror}):
\be
\begin{array}{ccccc}
Y = \MH (G,C) & \; & \; & \; & \MH (\LG,C) = \widetilde{Y} \\
\; & \searrow & \; & \swarrow & \; \\
\; & \; & \BB & \; & \;
\end{array}
\label{mhmhmirror}
\ee
This fibration is holomorphic\footnote{Note, that our conventions for $I,J,K$
and $\omega_I$, $\omega_J$, $\omega_K$ here agree with \cite{Hitchin} and differ from
\cite{GW} by a cyclic rotation of three complex structures $I \to J \to K \to I$.}
in complex structure $I$
and Lagrangian with respect to both $\omega_J$ and $\omega_K$,
so that $\BB$ and $\FF$ are of type $(B,A,A)$
in the terminology of section \ref{sec:hyperk}.

There is also a version of this story for Riemann surfaces with punctures,
which have $M = \M_{{\rm flat}} (G ,C; \frak C)$ as the classical phase space.
(To avoid cluttering, as in section \ref{sec:qart} we write most of the formulas
for a Riemann surface with a single puncture.)
As described in \eqref{mparabfibr}, this moduli space has the structure of
a symplectic fibration with the fiber $\frak C$ and symplectic form \eqref{omparabfibr}.
Much like $M = \M_{{\rm flat}} (G ,C; \frak C)$,
its natural complexification $Y = \M_{{\rm flat}} (G_{\C} ,C; \frak C_{\C})$
combines \eqref{yorbitcplx} and \eqref{yflatcplx} in a single
moduli space of flat $G_{\C}$ connections on $C$,
such that the holonomy of the connection around the puncture
takes values in a prescribed conjugacy class $\frak C_{\C}$.
Under mirror symmetry, this condition is replaced by a similar
condition, but for the Langlands dual group $\LG_{\C}$.

Namely, in this case, the mirror manifold $\tilde Y$
is the moduli space of semi-stable parabolic Higgs bundles on~$C$
with the structure group~$\LG$, {\it cf.} \eqref{mhmhmirror}.
In particular, it is a hyper-K\"ahler manifold and, in complex structure $J$,
can be identified with the moduli space $\M_{{\rm flat}} (\LG_{\C} ,C; \dual{\frak C}_{\C})$,
where $\dual{\frak C}_{\C} \subset \LG_{\C}$ is a complex conjugacy class
dual to $\frak C_{\C}$,
\be
\Phi_{{\rm mirror}} ~:~~~ \frak C_{\C} \; \rightarrow \; \dual{\frak C}_{\C} \,.
\label{conjcmirror}
\ee
This map is rather non-trivial.
It preserves the dimension of conjugacy classes,
as well as some other invariants described in \cite{Rigid}.

\example{$G=Sp(2N)$}
In section \ref{sec:qart}, we mentioned the minimal orbit of $B_N$.
For balance, now let us consider a group $G_{\C}$ of Cartan type $C_N$.
The minimal conjugacy class $\frak C_{{\rm min}}$,
{\it i.e.} the conjugacy class in $G_{\C}$ of the smallest dimension,
is the class of a unipotent element $U = \exp (u)$,
where $u_{ij} = \nu_i \nu_j$ is a rank-1 symmetric matrix.
It is parametrized by a vector $\nu$, defined up to a symmetry $\nu \to - \nu$,
so that
\be
\overline{\frak C}_{{\rm min}} \; \cong \; \C^{2N} / \Z_2 \,.
\ee
The dual conjugacy class $\dual{\frak C}_{{\rm min}} \subset \LG_{\C}$
of $B_N$ is the $2N$-dimensional conjugacy class of a semisimple element
\be
\dual{S} \; = \; \diag (+1, -1, -1, \ldots, -1) \,.
\ee
\endexample

In general, the holonomy $V \in G_{\C}$ can be written as $V = SU$,
where $S$ is semisimple, $U$ is unipotent, and $S$ commutes with $U$.
The duality map \eqref{conjcmirror} transforms
the semisimple and unipotent data in a non-trivial way.

\begin{conjecture}[\cite{Rigid}]
Let $\frak C_{\C}$ be a unipotent conjugacy class
(or a semisimple conjugacy class obtained by a deformation of $\frak C_{\C}$).
Then, the parameter $\frac{1}{2\pi} \log \dual{S}$ of the dual conjugacy class $\dual{\frak C}_{\C}$
is equal to the central character of (any) $G_{\R}$-representation
obtained by quantizing $\frak C_{\R} \subset \frak C_{\C}$.
\end{conjecture}



\subsection{Mirror of the Lagrangian brane $\CB'$}

Now let us introduce branes.
(For simplicity, we avoid punctures until section \ref{sec:punctorus}.)
In the $A$-model approach, quantization of $M$ is achieved by
studying the space of open strings (= space of morphisms)
between two $A$-branes, $\CB'$ and $\Bcc$, defined in section \ref{sec:amodel}.
The Lagrangian $A$-brane $\CB'$ is supported on $M \subset Y$,
while the coisotropic brane $\Bcc$ is supported on all of~$Y$
and carries a non-trivial Chan-Paton line bundle $\CL$ with curvature $F = {\rm Re}\, \Omega$.
Both of these branes turn out to be ``automatically'' compatible with
the hyper-K\"ahler structure on $Y$, thus, providing another
illustration of a phenomenon that we observed in some examples before.
Namely, $\CB'$ and $\Bcc$ both happen to be branes of type $(B,A,A)$.

In the case of $\CB'$ this follows from the fact that
$M = \M_{{\rm flat}} (G,C)$ is a component of the fixed point set
of the involution $\tau : Y \to Y$ that changes the sign of the Higgs field \cite{Hitchin}:
\be
\tau:~~ (A,\phi) \; \mapsto \; (A,-\phi) \,.
\label{mhtau}
\ee
This involution\footnote{Recall (from section \ref{sec:amodel}) that
the involution $\tau$ (antiholomorphic in complex structure $J$,
in which $\Omega$ is holomorphic) is needed for unitarity.}
is holomorphic in complex structure $I$ and antiholomorphic
in complex structures $J$ and $K$, so that $M$ is analytic in complex
structure $I$ and Lagrangian with respect to $\omega_J$ and $\omega_K$.
In the conventions \eqref{omegahk} - \eqref{fomhk}, it means that
$\CB'$ is not only a good $A$-brane in the $A$-model of $Y$ with $\omega_Y = \omega_K$,
but also can be viewed as a $B$-brane in the $B$-model of $(Y,I)$,
or else as an $A$-brane in the $A$-model of $(Y,\omega_J)$.
Using \eqref{fcois} and \eqref{bccfirstchern}
one can easily verify that the canonical coisotropic brane $\Bcc$
is also a brane of type $(B,A,A)$.

As pointed out in \eqref{hinbi},
we can take advantage of the fact that $\CB'$ and $\Bcc$ are compatible
with the hyper-K\"ahler structure on $Y$, and approach the problem from
the vantage point of the complex structure $I$, in which $\Bcc$ simply
corresponds to a holomorphic line bundle $\CL \to Y$
with the first Chern class $c_1 (\CL) = \omega_I$.
Then, the dimension of the Hilbert space $\CH$ can be computed
with the help of the Hirzebruch-Riemann-Roch formula similar to \eqref{GRRtheorem},
\be
\sum_i (-1)^i \dim H^i (M,\CL) \; = \; \int_M \ch (\CL) \wedge {\rm Td} (M) \,.
\label{HRRformula}
\ee
The spaces $H^i (M, \CL)$ are trivial for $i>0$,
so that \eqref{HRRformula} gives
\be
\dim \CH = \dim H^0 (M, \CL) = \int_M e^{\omega_I} \wedge {\rm Td} (M) \,,
\label{dimhmhini}
\ee
which, for $G = SU(N)$, indeed leads to the Verlinde formula \eqref{hdimforsun}.
However, our aim here is to approach the quantization of $M = \M_{{\rm flat}} (G,C)$
and, in particular, the calculation of $\dim \CH$ via mirror symmetry.\\

Mirror symmetry maps $(B,A,A)$ branes $\CB'$ and $\Bcc$ on $Y$
into $(B,B,B)$ branes $\tCB'$ and $\tBcc$ on $\tilde Y$.
In other words, $\tCB'$ and $\tBcc$ are good $B$-branes
with respect to all complex structures on the dual moduli space $\tilde Y = \MH (\LG,C)$.
Following our discussion in section \ref{sec:hyperk},
we expect that they correspond to hyperholomorphic bundles or sheaves on $\tilde Y$,
which (with a small abuse of notations) we also denote by $\tCB'$ and $\tBcc$.

To get an idea of what $\tCB'$ looks like,
it is instructive to start with a simple example of abelian gauge theory with gauge group $G = U(1)$.
In this case, the Hitchin fibration \eqref{mhmhmirror} is trivial, and
\be
Y \; = \; \BB \times \FF \,,
\label{mhtorus}
\ee
just like in our ``toy model'' considered in section \ref{sec:torus}.
In fact, the toy model of section \ref{sec:torus} arises as a special case
of the present discussion when $C$ is a Riemann surface of genus $g=1$.
More generally, for $G = U(1)$ we have
\be
\FF = {\rm Jac} (C)
\qquad , \qquad
\BB = H^0 (C;K_C)
\label{mhtorusfb}
\ee
where ${\rm Jac} (C)$ is the Jacobian of $C$ and $K_C$ is the canonical bundle.
Furthermore, in this case $\CB'$ is a Lagrangian brane supported on $M=\FF$.
(As noted earlier, the Hitchin fiber $\FF$ is always of type $(B,A,A)$,
and so is the brane $\CB' = \CB_{\FF}$.)
We already discussed the mirror transform of such branes in \eqref{syzfp}.
Indeed, dualizing the fiber $\FF$ we obtain a mirror brane
\be
\tCB' = \CB_p
\label{zerobrane}
\ee
supported at a point $p \in \widetilde{Y} = \BB \times \widetilde{\FF}$.
Clearly, this is a brane of type $(B,B,B)$;
in any $B$-model of $\tilde Y$ ({\it i.e.} for any complex structure on $\tilde Y$)
it corresponds to the skyscraper sheaf $\CO_p \in \CD^b (\tilde Y)$.

When the gauge group $G$ is non-abelian, the mirror brane $\tCB'$ is also
a $0$-brane, in a sense that its support is a point on $\widetilde{Y} = \MH (\LG,C)$,
but now it has a non-trivial ``inner structure.''
Specifically, the $(B,B,B)$ brane $\tCB'$ is supported
at the ``most singular point'' $(A,\phi) = (0,0)$ on $\MH (\LG,C)$,
with a pole for the ${}^L{\g}_{\C}$-valued fields $\sigma$ and $\bar \sigma$
that corresponds to the principal (a.k.a. regular) $\mathfrak{su} (2)$ embedding~\cite{WittenNahm,FG}:
\be
\rho_{{\rm princ}} :~~ \mathfrak{su}(2) \; \to \; {}^L{\g}
\label{princemb}
\ee
In particular, the mirror of the Lagrangian brane $\CB'$ supported
on $M = \M_{{\rm flat}} (G,C)$ does not admit a simple geometric
description in the $B$-model of $\widetilde{Y} = \MH (\LG,C)$,
roughly speaking, because
all the information about $M$ is now clumped at the ``most singular point'' of $\widetilde{Y}$.
In order to give a proper description of the $(B,B,B)$ brane $\tCB'$
one needs either to introduce ramification (as in section \ref{sec:punctorus} below)
or to extend the $B$-model of $\widetilde{Y}$
to account for the fields $\sigma$ and $\bar \sigma$.
We will not attempt to formulate such a description here and, instead,
focus on those $(B,B,B)$ branes on $\widetilde{Y}$
that can be described in the language of hyperholomorphic bundles or sheaves.

In fact, we expect $\tBcc$ to be a nice example of a $(B,B,B)$ brane
that corresponds to a hyperholomorphic bundle on $\widetilde{Y} = \MH (\LG,C)$.
As for the Lagrangian brane $\CB'$, we can consider close cousins
of the brane supported on the moduli space of flat connections,
$M = \M_{{\rm flat}} (G,C)$. Namely, we can take $\CB'$ to be
a Lagrangian brane supported on another component
of the fixed point set of the involution \eqref{mhtau}.
Clearly, all such branes are automatically of type $(B,A,A)$,
and some of them even admit a nice geometric interpretation
as branes supported on $M = \M_{{\rm flat}} (G_{\R},C)$,
for other real forms $G_{\R}$ of the complex group $G_{\C}$ \cite{Hitchin}.
As reviewed in \eqref{abholonomies} - \eqref{omcs},
these moduli spaces provide an excellent laboratory for
the quantization problem, with many applications to gauge theory.


\subsection{Mirror of the coisotropic $(B,A,A)$ brane $\Bcc$}

Now let us consider the mirror transform of the canonical coisotropic $(B,A,A)$ brane $\Bcc$
with a Chan-Paton bundle of curvature $F = \omega_I$.
Among all coisotropic branes on $Y \cong \MH (G,C)$,
this brane has a number of special properties.
In order to describe them in detail, let us introduce a complexified K\"ahler form
\be
\omega_I + i B \; = \; \frac{1}{\hbar} \omega_*
\label{mhh2generator}
\ee
where $\omega_*$ is the image in de Rham cohomology of a generator of $H^2 (Y,\Z) \cong \Z$.
One peculiar property of the $(B,A,A)$ brane $\Bcc$ is that it exists (with $B=0$)
only for discrete values of $\hbar$:
\be
\hbar \; = \; \frac{1}{k} \,,
\label{hviak}
\ee
where $k \in \Z$ is the ``level.''
This agrees well with the fact \eqref{khinteger} that
$M = \M_{{\rm flat}} (G,C)$ should be quantizable precisely for these values of $\hbar$.
Mirror symmetry \eqref{mhmhmirror} maps $\Bcc$ to $\tBcc$ and acts on the parameter $\hbar$
as\footnote{The setup
of the present section arises in topological gauge theory on a 4-manifold $\R^2 \times C$.
In that context, the complex parameter $\hbar$
is identified with the coupling constant of the four-dimensional gauge theory \cite{KW}.}
\be
\hbar \; \to \; {}^L{\hbar} = - \frac{1}{\hbar}
\label{hdual}
\ee
where, for simplicity, we assumed that $\g$ is simply laced.
It follows that, just like $\Bcc$, its mirror $\tBcc$ exists only
for a discrete set of values of $\hbar$, namely
\be
{}^L{\hbar} \; = \; - k \,.
\ee
How and why this happens can be seen already in a basic example of abelian
gauge theory (which, in the special case of $g=1$, was discussed in detail
in section \ref{sec:torus}, and has a straightforward generalization to
arbitrary genus, based on \eqref{mhtorus} - \eqref{mhtorusfb}).

To learn more about the brane $\tBcc$ and about the $B$-model approach to
quantization of $M = \M_{{\rm flat}} (G,C)$ we can examine the problem from
the viewpoint of complex structure $I$.
As explained in section \ref{sec:hyperk}, in the $B$-model of $(Y,I)$
the original brane $\Bcc$ corresponds to a complex line bundle $\CL \to Y$
with the first Chern class \eqref{bccfirstchern}, {\it i.e.} with the Chern character
\be
\ch (\Bcc) \; = \; \exp (\omega_I) \,.
\ee
Moreover, in complex structure $I$ mirror symmetry \eqref{mirrormap}
is simply the Fourier-Mukai transform \eqref{fmmap}, and \eqref{chmap}
gives the Chern character of the dual brane $\tBcc$.
Although in general $\tBcc$ turns out to be a higher rank $(B,B,B)$ brane,
we expect that it corresponds to an ordinary hyperholomorphic bundle
on $\tilde Y$ (as opposed to a more complicated object in $\CD^b (\tilde Y)$)
with the Chern character $\ch (\tBcc)$.

For concreteness, let us take $G=SU(2)$. (We will try, however,
to focus on general properties of $\Bcc$ that will have a clear analog for other groups.)
Then, even without getting too much into details of the geometry of $\MH (\LG,C)$,
we can say that the answer for $\ch (\tBcc)$ must have the following structure,
familiar from \eqref{chbccmirrortoy} and \eqref{chbccmirrork3}:
\be
\ch (\tBcc) \; = \; \frac{2^g}{\hbar^{3g-3}} + \; \ldots \; - \BB \,.
\label{chbccmirrormh}
\ee
Here, the first term (of degree $0$) is the well-known expression \cite{BNR}
for the volume of the Hitchin fiber, ${\rm Vol} (\FF) = \int_{\FF} e^{\omega_I}$,
which according to \eqref{tbccrank} determines the rank of the mirror
of the coisotropic brane $\Bcc$.
The last term (to be discussed shortly) does not depend on $\hbar$,
while the remaining terms (denoted by ellipsis) all appear with
{\it negative} powers of $\hbar$.
In particular, all the terms in \eqref{chbccmirrormh}, except
for the last one, vanish in the ``extreme quantum'' limit:
\be
\hbar \; \to \; \infty \,.
\label{veryquantum}
\ee
There is a simple explanation for this.
Indeed, in the limit \eqref{veryquantum} the curvature $F = \omega_I$
of the Chan-Paton bundle of $\Bcc$ goes to zero,
and $\Bcc$ becomes a rank-1 brane supported on all of $Y$ with a trivial Chan-Paton bundle.
As an object of the derived category $\CD^b (Y,I)$ this limiting brane is simply
the structure sheaf $\CO_Y$.
Even though it preserves different supersymmetry, $(B,B,B)$ instead of $(B,A,A)$,
the brane $\CB = \CO_Y$ can help us to understand what happens to the leading term in
$\ch (\Bcc) =  1 + O (\hbar^{-1})$ under the mirror map.
Indeed, the mirror of $\CB = \CO_Y$ is a $(B,A,A)$ brane $\tCB$ on $\tilde Y$
supported on a section of the dual Hitchin fibration \eqref{mhmhmirror},
whose homology class accounts for the last term in \eqref{chbccmirrormh}.

If we naively try to compute the dimension of the Hilbert space $\CH$ from
the partial answer \eqref{chbccmirrormh}, as we did {\it e.g.} in \eqref{torusbmodel},
we obtain an expression that looks like
\be
\dim \CH \; = \; {\rm vol} (\FF) \cdot k^{3g-3} + \ldots + 1 \,,
\label{dimhfirstguess}
\ee
where the first (resp. last) term comes from the corresponding term in \eqref{chbccmirrormh}.
To make the meaning of these terms more transparent, we used \eqref{hviak}
to replace $\hbar$ by $k$, and wrote ${\rm vol} (\FF)$ for the volume of the fiber $\FF$
with respect to the normalized symplectic form $\omega_*$ introduced in~\eqref{mhh2generator}.
Even though our derivation of \eqref{dimhfirstguess} was somewhat heuristic
since we treated $\tCB'$ as an ordinary zero-brane \eqref{zerobrane} ignoring the pole \eqref{princemb},
the answer \eqref{dimhfirstguess} does capture correctly certain aspects of the Verlinde formula.

For example, it is clear that \eqref{dimhfirstguess} is a polynomial in $k$
(because $\ch (\tBcc)$ is a polynomial in $\hbar^{-1}$).
Furthermore, the last term in \eqref{dimhfirstguess} is the constant term of this polynomial.
This follows from the fact that the last term of \eqref{chbccmirrormh}
is the only term in $\ch (\tBcc)$ independent of $\hbar$.
This agrees well with the behavior of the Verlinde formula
in the extreme quantum limit \eqref{veryquantum}:
\be
\dim \CH \;\overset{k \to 0}{\longrightarrow}\; 1 \,,
\ee
valid for any genus $g$, {\it cf.} \eqref{dimhgenus2} for $g=2$.
However, as that genus-$2$ example also illustrates, even though the polynomial
\eqref{dimhfirstguess} has the correct degree, the coefficient of the leading
term (for large $k$) is {\it not} what we expect it to be.
Indeed, according to \eqref{hdimvolm}, the (coefficient of the) leading term
in the Verlinde formula should be the volume of $M = \M_{{\rm flat}} (G,C)$, not the volume of $\FF$.
This is also clear from the viewpoint \eqref{dimhmhini} of the $B$-model of $(Y,I)$.

The volumes of $M$ and $\FF$ are quite different.
For example, for $G=SU(2)$ the normalized volume of $M$,
computed with respect to the symplectic form $\omega_*$,
is given by the following formula \cite{W2}:
\be
{\rm vol} (M) \; = \; \frac{2 \cdot \zeta (2g-2)}{(2 \pi^2)^{g-1}}
\label{volumeM0}
\ee
which is quite different from a (much simpler) expression for the volume of $\FF$.
The first few values of \eqref{volumeM0} are listed in Table \ref{tab:volumes}.

\begin{table}
\begin{center}
\begin{tabular}{c|c|c}
~~genus~~&$\phantom{\oint} {\rm vol} (M) \phantom{\oint}$&$\phantom{\oint} {\rm vol} (\FF)$~~\\
\hline
$g=2$&$\frac{1}{6}$&$4$\\
$g=3$&$\frac{1}{180}$&$8$\\
$g=4$&$\frac{1}{3780}$&$16$\\
$g=5$&$\frac{1}{75600}$&$32$\\
$\ldots$&$\ldots$&$\ldots$\\
\end{tabular}
\end{center}
\begin{caption}\noindent\small{Listed here are the volumes of the moduli space $M$
and the Hitchin fiber~$\FF$ for $G=SU(2)$ and small values of $g$,
computed with respect to the symplectic form~$\omega_*$.}
\label{tab:volumes}
\end{caption}
\end{table}

The fact that \eqref{dimhfirstguess} correctly captures the constant term of the Verlinde formula,
and not the top-degree term, should not be surprising. After all, \eqref{chbccmirrormh}
correctly captures the part of $\ch (\tBcc)$ that does not depend on $\hbar$,
whereas there are several terms in $\ch (\tBcc)$ --- suppressed in \eqref{chbccmirrormh} --- that
contribute to the leading behavior of $\dim \CH \sim k^{3g-3}$.
The number of such terms grows quickly with the genus $g$ of the Riemann surface $C$.


\section{The Verlinde formula via mirror symmetry}
\label{sec:punctorus}

\hfill{\vbox{\hbox{\it The more success the quantum theory has}
\hbox{\it the sillier it looks.}}}

\hfill{\vbox{\hbox{Albert Einstein}}}

\medskip


Now, let us analyze the space $\CH$ of section \ref{sec:cstheory}
and its dimension more carefully.
In particular, we wish to see how various terms in the Verlinde formula
\eqref{verlindegeneral} arise in the $B$-model approach,
where the target space is the moduli space of Higgs bundles on $C$
with the structure group~$\LG$.

To keep our discussion concrete and simple at the same time,
we take $G=SU(2)$ and focus on a specific example,
to which we secretely prepared ourselves in section \ref{sec:bmodel}.
Namely, we take $C$ to be a torus with a single puncture,
around which the gauge field has a holonomy \eqref{su2holonomy} labeled by a weight $\lambda$.
Then, much like in examples considered in section \ref{sec:bmodel},
we have $\dim_{\R} M = \dim_{\C} Y = 2$
and the Verlinde formula \eqref{verlindeparabolic} gives:
\be
\dim \CH \; = \; k - \lambda + 1
\label{verlindetorus}
\ee
for even values of $\lambda$ and sufficiently large $k$.
In other words, in this case the Verlinde formula
is a simple polynomial of degree $n=1$
with only two non-trivial coefficients,
$a_0$ and $a_1$ in the notations of \eqref{verlindegeneral}.
Nevertheless, understanding these coefficients via branes
will be an illuminating and enjoyable exercise.

First, let us summarize the relevant geometry of the space
$M = \M_{{\rm flat}} (G,C)$, its complexification $Y = \M_{{\rm flat}} (G_{\C},C)$,
and the mirror $\tilde Y = \M_{{\rm flat}} (\LG_{\C},C)$,
{\it cf.} \eqref{yflatcplx} and \eqref{mhmhmirror}.
Since the fundamental group of a 2-torus with one puncture is a free group of rank two:
\be
\pi_1 (C) \; = \; \langle a,b,c ~\vert~ ab a^{-1} b^{-1} = c \rangle \,,
\ee
its $SU(2)$ and $SL(2,\C)$ character varieties $M$ and $Y$
admit a very simple and explicit description \eqref{parabhol}.
For instance, the space $Y$ of homomorphisms $\rho : \pi_1 (C) \to SL(2,\C)$
with a suitable boundary condition \eqref{su2holonomy} at the puncture
can be described rather explicitly as an affine hypersurface in $\C^3$ with coordinates
\begin{eqnarray}
x & = & \tr (\rho (a)) \nonumber \\
y & = & \tr (\rho (b)) \nonumber \\
z & = & \tr (\rho (ab)) \nonumber
\end{eqnarray}
defined by
\be
Y~: \qquad
x^2 + y^2 + z^2 + xyz \; = \; \tr V + 2 \,.
\label{toruscubic}
\ee
For a surface $Y$ defined by the zero locus of a polynomial $f(x,y,z)$
the holomorphic form \eqref{omegahk} can be written as
\be
\Omega = \frac{1}{4 \pi^2 \hbar} \; \frac{dx \wedge dy}{\partial f / \partial z}
= \frac{1}{4 \pi^2 \hbar} \; \frac{dx \wedge dy}{xy + 2z} \,.
\label{cubicholom}
\ee
When $\tr V = 2$ ({\it i.e.} $\alpha = 0$), we obtain a cubic surface with
four simple singularities of type $A_1$ (double points) at
\be
(-2,-2,-2) \,, \quad (-2,2,2) \,, \quad (2,2,-2) \,, \quad {\rm and} \quad (2,-2,2) \,.
\ee
This singular surface, called the Cayley cubic,
is simply a $\Z_2$ quotient of $\C^* \times \C^*$:
\be
Y \; = \; ( \C^* \times \C^* ) / \Z_2 \,.
\label{sl2ctorus}
\ee
A more direct way to see this is to note that,
for the special value of the holonomy parameter $\alpha =0$,
we have $V = {\bf 1}$ and the defining equation \eqref{parabhol}
reduces to that of a 2-torus without punctures, {\it cf.} \eqref{abholonomies}.
On the other hand, since the fundamental group of a torus
is abelian, $\pi_1 (T^2) = \Z \times \Z$, the holonomies
of the complexified gauge connection $\CA = A + i \phi$
around the $A$- and $B$-cycles of $T^2$ can be
simultaneously conjugated to a maximal torus $\mathbb{T}_{\C} \subset G_{\C}$.
Hence, $\M_{{\rm flat}} (G_{\C},T^2) = (\mathbb{T}_{\C} \times \mathbb{T}_{\C}) / \CW$
where $\CW$ is the Weyl group. In the present case, this gives \eqref{sl2ctorus}
because $\mathbb{T}_{\C} = \C^*$ and $\CW = \Z_2$.

Now, it is clear that, for $\alpha =0$, the space $Y$ in
the present example is simply a $\Z_2$ quotient of that in section \ref{sec:torus}:
\be
Y \; = \; ( T^2 \times \R^2 ) / \Z_2 \,.
\label{t2z2quotient}
\ee
Moreover, the real slice $M = T^2 / \Z_2$, sometimes called the ``pillow case,''
is the moduli space of flat $SU(2)$ connections on the (punctured) torus.
Turning on the holonomy parameter $\alpha$ removes the four $\Z_2$ singularities
and deforms \eqref{t2z2quotient} into a smooth complex surface \eqref{toruscubic}.
This is the viewpoint of complex structure $J$, in which $Y$ is identified
with the moduli space of flat $SL(2,\C)$ connections on $C$.

Since we are interested in branes of type $(B,A,A)$
let us consider what happens in complex structure $I$,
in which $M \subset Y$ is holomorphic and $Y$ is naturally identified with the moduli space
of semi-stable parabolic Higgs bundles on $C$, {\it cf.} \eqref{ymhhiggs}.
In complex structure $I$, the parameter $\alpha$ is a K\"ahler structure parameter.
For $\alpha \ne 0$ the four $\C^2 / \Z_2$ orbifold singularities of $Y$
are resolved, and we denote by $D_i$, $i=1, \ldots, 4$, the corresponding exceptional divisors.
As a result, $Y$ has homology (for generic values of $\alpha$):
\be
H_2 (Y) \; \cong \; \Z^5 \,.
\ee
One can also deform the $\C^2 / \Z_2$ singularities by turning on
a complex structure parameter $\beta + i \gamma$, which in the present
example corresponds to introducing a pole for the Higgs field $\phi$
at the puncture $p \in C$.
This leads to a closely related model, recently considered in \cite{FW}.
Since such complex structure deformations create exceptional cycles
which are not holomorphic\footnote{In general, the hyper-K\"ahler metric
on $Y$ depends on a triple of ``moment maps'' $(\alpha,\beta,\gamma)$,
such that for generic values of these parameters the exceptional cycles
are holomorphic in complex structure
$$
\CI = \frac{\alpha I + \beta J + \gamma K}{\sqrt{\alpha^2 + \beta^2 + \gamma^2}} \,.
$$}
in complex structure $I$, we shall mainly focus on
the situation with $\alpha \ne 0$ and $\beta = \gamma = 0$.

In complex structure $I$, the surface $Y$ has the structure
of the elliptic fibration~\eqref{mhmhmirror}.
Indeed, if $z$ and $w$ are complex coordinates on $T^2$ and $\R^2$ in \eqref{t2z2quotient},
then there is a map $\pi : Y \to \BB$, sending $(z,w) \mapsto b := w^2$
and exceptional divisors to zero.
The generic fibers of this map are $\FF \cong T^2$
and the only singular fiber is the ``nilpotent cone'' $\CN := \pi^{-1} (0)$,
which in the present case has five irreducible components (all rational):
\be
\CN \; = \; M \cup \bigcup_{i=1}^{4} D_i \,.
\label{nilconetoy}
\ee
The homology classes of $M$ and $D_i$ are independent and generate $H_2 (Y)$.
A quick look at the intersection numbers shows that $Y$
is indeed an elliptic fibration with one singular fiber over $b=0$
of Kodaira type $I_0^*$, {\it i.e.} with the intersection form $\tilde D_4$
(in the basis $\{ M, D_1, \ldots, D_4 \}$):
\be
\begin{pmatrix}
-2 & 1 & 1 & 1 & 1 \\
1 & -2 & 0 & 0 & 0 \\
1 & 0 & -2 & 0 & 0 \\
1 & 0 & 0 & -2 & 0 \\
1 & 0 & 0 & 0 & -2
\end{pmatrix}
\label{torusintersection}
\ee
This intersection matrix has only one null vector, which therefore
must be identified with the class of the elliptic fiber,
\be
[\FF] \; = \; 2 [M] + \sum_{i=1}^{4} [D_i] \,,
\label{fmhomoltoy}
\ee
and implies the following relation among the volumes:
\be
{\rm Vol} (\FF) \; = \; 2 {\rm Vol} (M) + \sum_{i=1}^{4} {\rm Vol} (D_i) \,.
\label{volsumtoy}
\ee
Indeed, when $\alpha=0$ we have ${\rm Vol} (D_i)=0$ and this equation
simply expresses the fact that $\FF$ is a double cover of $M$,
which is clear from the $\Z_2$ quotient \eqref{t2z2quotient}.
Apart from this multiplicity factor (due to a singularity),
the relation \eqref{fmhomoltoy} is the familiar statement
that different fibers of $\pi : Y \to \BB$ are homologous.

At this stage, we have everything we need to verify the Verlinde formula \eqref{verlindetorus}.
In the $A$-model of $Y$, $\CB'$ is a Lagrangian brane supported on $M \subset Y$
and $\Bcc$ is a coisotropic brane with a Chan-Paton bundle $\CL$ of curvature $F = \omega_I$.
Away form the singular fiber over $b=0$, the brane $\Bcc$ (resp. its dual $\tBcc$)
is essentially the same as the one considered in section \ref{sec:torus}.
(For simplicity, one can keep in mind the special case $\alpha=0$,
for which $Y$ is given by the $\Z_2$ quotient \eqref{t2z2quotient}.)
The only important effect of the $\Z_2$ quotient is that $\FF$ is a double cover of $M$
and $\hbar = \frac{1}{2k}$, so that
\be
{\rm Vol} (\FF) = \int_{\FF} \omega_I = 2k \,,
\label{fvoltorus}
\ee
This relation actually holds for all values of $\alpha$,
as can be easily verified by a direct evaluation of
the period integral of the holomorphic $2$-form \eqref{cubicholom}:
\be
\int_{\FF} \Omega \; = \; \frac{1}{(2 \pi i)^3 \hbar}
\iiint_{|x|=|y|=|z|=1} \frac{dx \wedge dy \wedge dz}{f(x,y,z)}
\; = \; \frac{1}{\hbar} \,.
\label{fperiod}
\ee
Similarly, we find\footnote{Notice, when $\alpha =0$ we have
${\rm Vol} (\FF) = 2 {\rm Vol} (M)$, in agreement with \eqref{volsumtoy}.}
\be
\int_{M} \Omega \; = \; \frac{1}{\hbar} \left( \frac{1}{2} - \alpha \right)
\; = \; k - \lambda \,,
\label{mperiod}
\ee
where $\alpha = \frac{\lambda}{2k}$ was used in the last equality,
{\it cf.} \eqref{su2holonomy}.
Then, this gives us the correct answer for the dimension of $\CH$,
consistent with the Verlinde formula \eqref{verlindetorus},
\be
\dim \CH \; = \; {\rm Vol} (M) + 1 \; = \; k - \lambda + 1 \,,
\ee
where we also used \eqref{dimhmhini} and ${\rm Td} (M) = e^{c_1 (M)/2} \widehat{A} (M)$.

Note, according to the general formula \eqref{tbccrank}, the volume
of the elliptic fiber \eqref{fvoltorus} determines the rank of the mirror
of the coisotropic brane $\Bcc$,
\be
\rank (\tBcc) \; = \; 2k \,.
\label{bccrankpunct}
\ee
This is our first hint that the mirror of the Lagrangian brane $\CB'$
should be a ``fractional brane.'' Indeed, if $\tCB'$ were
a regular zero-brane on $\tilde Y$ represented by
a skyscraper sheaf $\CO_p \in \CD^b (\tilde Y)$, as in \eqref{zerobrane},
it would contribute $2k$ (instead of $k$) to the dimension of $\CH$, {\it cf.} \eqref{torusbmodel}.
Therefore, we expect that $\tCB'$ should be roughly a ``half'' of the ordinary
zero-brane $\CB_p$ supported at a generic point $p \in \tilde Y$.
As we shall see below, this is indeed the case.

Before we proceed, let us remark that
the relations \eqref{nilconetoy} - \eqref{volsumtoy} have
analogs for more general moduli spaces of (parabolic) Higgs bundles.
For example, for $SU(2)$ Higgs bundles on a Riemann surface of genus $g$
the nilpotent cone 
has $g$ irreducible components, each of complex dimension $3g-3$ \cite{Hitchin,Thaddeus}:
\be
\CN \; = \; M \cup \bigcup_{i=1}^{g-1} D_i \,.
\label{nilcone}
\ee
Here, $M$ is the classical phase space of $SU(2)$ Chern-Simons gauge theory on $C$,
and each $D_i$ is the locus of those stable Higgs bundles $\CE = E  \;\overset{\phi}{\to}\; E \otimes K_C$
which have a unique subbundle $\CL_i$ of degree $(1-i)$ killed by the non-zero Higgs field~$\phi$.
Moreover, in this case,
the middle dimensional homology $H_{6g-6} (Y)$ has dimension $g$
and is freely generated by the homology classes of irreducible
components of the nilpotent cone~\eqref{nilcone}.
Similarly, the analog of \eqref{fmhomoltoy} is the following
relation (due to T.~Hausel),
\be
[\FF] \; = \; \sum_i 2^{\dim (F_i)} [D_i] \,,
\ee
where $F_i$ are the connected components of the fixed point set
of the circle action $(A,\phi) \to (A, e^{i \xi} \phi)$,
and $D_i$ are the corresponding components of the nilpotent cone (with $D_0 \equiv M$),
see \cite{Hitchin,Thaddeus,Hausel,HT} for further details.

Returning to our basic example of a genus $1$ curve $C$ with one puncture,
let us consider the $B$-model of the mirror variety $\tilde Y$ which,
according to \eqref{mhmhmirror}, we identify with the moduli space
of parabolic Higgs bundles for the Langlands dual group $\LG = SO(3)$.
Much like $Y$ itself, its mirror $\tilde Y$ is an elliptic fibration
$\tilde \pi : \tilde Y \to \BB$,
with generic fibers $\tilde \FF_b = H^1 (\FF_b, U(1)) \cong T^2$
and one singular fiber over $b=0$.
Since in general mirror symmetry for Calabi-Yau 2-folds
is believed to preserve the Kodaira type of singular fibers,
we expect that $\tilde Y$ also has a singular fiber of type $I_0^*$ over $b=0$.
(Of course, the singularities of $\tilde Y$ may be only partially resolved
since mirror symmetry exchanges complex and symplectic structures.)
In order to show that this is indeed a correct guess,
in the present example it is convenient to construct
the moduli space $\tilde Y = \MH (\LG,C)$ as a quotient of $Y = \MH (G,C)$,
\be
\tilde Y \; \cong \; Y / \, \Xi \,,
\ee
which follows from the well-known isomorphism $SO(3) \cong SU(2) / \Z_2$.
Here, $\Xi = \Z_2 \times \Z_2$ is the ``group of sign changes''
generated by twists of the underlying gauge bundle $E \to C$ by line bundles of order 2.
The elements of this group act on the $SL(2,\C)$ character variety \eqref{toruscubic}
by reflections $(x,y,z) \mapsto (\pm x, \pm y, \pm z)$ with an even number of minus signs, see {\it e.g.} \cite{Goldman}.
The resulting quotient $\tilde Y = Y / \Xi$ is an elliptic surface
with three $\C^2 / \Z_2$ orbifold singularities located at $(x,y,z) = (\sqrt{2 + \tr V},0,0)$
and two other points obtained by permutations of $x$, $y$, and $z$.
All of these points lie on the singular fiber of $\tilde \pi : \tilde Y \to \BB$,
namely on the zero fiber~$\tilde \CN = \tilde \pi^{-1} (0)$.

\begin{figure}[htb]
\centering
\includegraphics[width=4in]{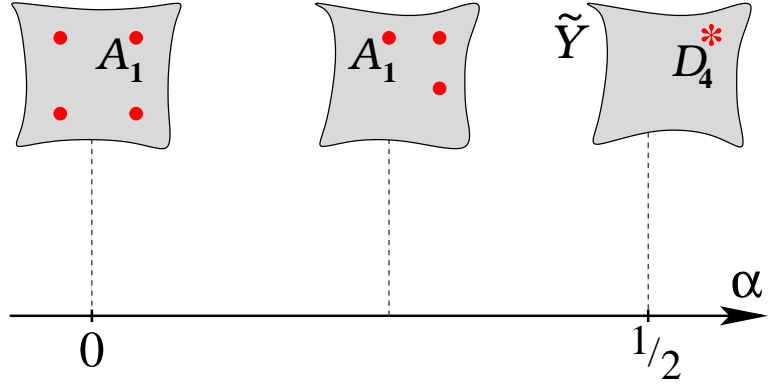}
\caption{Singularities of $\tilde Y$:
$a)$ four simple singularities of type $A_1$ when $\alpha = 0$,
$b)$ one singularity of type $D_4$ when $\alpha = \frac{1}{2}$, and
$c)$ three $A_1$ singularities for all other values $0 < \alpha < \frac{1}{2}$.}
\label{fig:singpts}
\end{figure}

In complex structure $\tilde J$, the singular surface $\tilde Y$ can be represented
as a zero locus of a cubic in $\C^3$, similar to \eqref{toruscubic},
\be
\tilde Y~: \qquad
x^2 + y^2 + z^2 + xyz \; = \; 2a^2 (x+y+z) + (4 - 4a^2 - a^4) \,,
\label{mirrorcubic}
\ee
where $a^2 = 2 - \tr V$.
Branes on this particular family of singular cubic surfaces
were studied in a closely related context in \cite{GukovRTN}.
In the new coordinates, the orbifold singularities of $\tilde Y$
are located at $(x,y,z) = (- \tr V,2,2)$
and two other points obtained by permutations of $x$, $y$, and $z$.
Notice, when $\tr V = - 2$ all three simple singularities of type $A_1$ collide and,
in fact, $\tilde Y$ develops a worse singularity of type $D_4$ at $(x,y,z) = (2,2,2)$,
as drawn schematically in Figure \ref{fig:singpts}.

On the other hand, when $\alpha = 0$ ({\it i.e.} $a = 0$)
the mirror geometry \eqref{mirrorcubic}
develops the fourth $A_1$ singularity at $(x,y,z) = (-2,-2,-2)$.
At this special value of $\alpha$ both $Y$ and $\tilde Y$
take the form of the Cayley cubic \eqref{sl2ctorus},
with possible values of the $B$-field equal to $0$ or $\frac{1}{2}$
in the direction of each exceptional divisor~\cite{Aspinwall}.
Which values are realized in our problem can be easily determined
via the connection with~\cite{GukovRTN}, where the same sigma-model
played an important role in the gauge theory approach to knot homologies.
Thus, in order to understand the basic operations in
knot theory (the skein relations) one needs to study
the special case of a four-punctured sphere $\cp^1 \setminus \{ p_1, p_2, p_3, p_4 \}$,
and the family of cubic surfaces \eqref{mirrorcubic} is precisely the moduli space
of flat $SL(2,\C)$ connections on a four-punctured sphere with holonomies $V_i$,
such that $\tr V_i = a$ for all $i=1, \ldots, 4$.

Likewise, the original cubic surface \eqref{toruscubic} incidentally is
the moduli space of flat $SL(2,\C)$ connections on $\cp^1 \setminus \{ p_1, p_2, p_3, p_4 \}$
with holonomy parameters
\be
\alpha_1 = \frac{\alpha}{4} - \frac{1}{2}
\qquad {\rm and} \qquad
\alpha_2 = \alpha_3 = \alpha_4 = \frac{\alpha}{4} \,,
\label{aaaa}
\ee
where we labeled each holonomy $V_i$ by a parameter $\alpha_i$
as in \eqref{su2holonomy}.
This fact can be used to determine the value of the $B$-field in
the mirror $B$-model of $\tilde Y$.
Indeed, the duality maps each holonomy parameter $\alpha_i$
to the ``quantum'' parameter $\eta_i$ associated with the $i$-th puncture \cite{Ramified},
\be
\Phi_{{\rm mirror}} ~:~~~ \alpha_i \; \rightarrow \; \eta_i \,.
\label{aetamirror}
\ee
In the mirror $B$-model with the target space $\tilde Y$,
the quantum parameters $\eta_i$ describe the flux of the $B$-field
through the corresponding $2$-cycles $\tilde D_i$.
This, together with \eqref{aaaa}, determines the value of the $B$-field:
\begin{eqnarray}
B & =: & \sum_{i=1}^4 \eta_i \tilde D_i  \label{etet} \\
& = & \left( \frac{\alpha}{4} - \frac{1}{2} \right) \tilde D_1
+ \frac{\alpha}{4} \tilde D_2 + \frac{\alpha}{4} \tilde D_3 + \frac{\alpha}{4} \tilde D_4 \,. \nonumber
\end{eqnarray}
The non-trivial $B$-field has an important effect in the $B$-model of $\tilde Y$.
In particular, one should remember that, in the presence of a $B$-field,
the Chern character $\ch (\tCB)$
always appears in a gauge invariant combination $e^{-B} \ch (\tCB)$
and the charge vector of a brane $\tCB$ is given by the modified
Mukai vector,
\be
v (\CB) = e^{-B} \ch (\CB) \sqrt{{\rm Td} (\tilde Y)} \,,
\label{bmukaicharge}
\ee
instead of \eqref{mukaicharge}.

As the parameter $\alpha$ gradually varies from $0$ to $\frac{1}{2}$,
the geometry of $\tilde Y$ interpolates between the two extreme
cases depicted in Figure \ref{fig:singpts},
so that three $A_1$ singularities remain unresolved.
With our choice of conventions, the corresponding exceptional divisors
are $\tilde D_2$, $\tilde D_3$, and $\tilde D_4$.
In other words, the 2-cycles $\tilde D_2$, $\tilde D_3$, and $\tilde D_4$ have zero volume
with respect to all K\"ahler forms on $\tilde Y$, whereas ${\rm Vol} (\tilde M)$ and
${\rm Vol} (\tilde D_1)$ vary with $\alpha$ in such a way that
their linear combination \eqref{volsumtoy} gives the volume of the elliptic fiber,
\be
{\rm Vol} (\tilde \FF) \; = \; 2 {\rm Vol} (\tilde M) + {\rm Vol} (\tilde D_1) \,,
\label{fdmrel}
\ee
and remains constant (independent of $\alpha$).
Indeed, for the cubic surface \eqref{mirrorcubic} with real values of $a$,
only $\tilde \omega_I$ has non-zero periods over these 2-cycles,
and an easy computation analogous to \eqref{fperiod} - \eqref{mperiod}
shows that, besides \eqref{fdmrel}, the periods obey the following relation,
\be
\int_{\tilde D_1} \tilde \omega_I \; = \; 2 \alpha \int_{\tilde \FF} \tilde \omega_I \,,
\ee
which will be useful to us below.
Note, in particular, that at $\alpha=0$ we have ${\rm Vol} (\tilde D_1) = 0$,
whereas at $\alpha = \frac{1}{2}$ the volume of $\tilde M$ vanishes.

Now, let us discuss $(B,B,B)$ branes on $\tilde Y$,
in particular, the branes $\tCB'$ and $\tBcc$ which are of major importance
in the $B$-model approach to quantization of $M = \M_{{\rm flat}} (G,C)$.
Starting with \eqref{syzfp}, by now we encountered several times
one particular $(B,B,B)$ brane, namely a regular zero-brane $\CB_p$,
which is dual to a $(B,A,A)$ brane $\CB_{\FF}$ supported
on a generic fiber of $\pi : Y \to \BB$.

In addition, the category of $B$-branes on $\tilde Y$ contains
``fractional'' zero-branes supported at the orbifold singularities of $\tilde Y$.
Specifically, the spectrum of branes at the Kleinian quotient singularity $\C^2 / \Gamma$
by a finite group $\Gamma \subset SL(2,\C)$ is described by $\CD^b_{\Gamma} (\C^2)$,
and fractional branes correspond to the simple objects of this category:
\be
\CB_i \; = \; \varrho_i \otimes \CO_p \,.
\label{sbranes}
\ee
Here, $\varrho_i$ are irreducible representations\footnote{Our
conventions are such that $\varrho_1$ always denotes the trivial representation.}
of $\Gamma$ and $\CO_p$ is the skyscraper sheaf supported at the origin of $\C^2$.
The category of fractional branes is equipped with an action
of the tensor category ${\rm Rep} (\Gamma)$.
For example, if $\Gamma = \Z_2$, as in our model with $0 \le \alpha < \frac{1}{2}$,
then at every orbifold point
there are two fractional branes $\CB_+$ and $\CB_-$ of charge $v (\CB_{\pm}) = (0,\pm 1, \frac{1}{2})$,
permuted by the sign representation of $\Gamma = \Z_2$
and left invariant by the action of the trivial representation
of $\Gamma = \Z_2$, {\it cf.} \cite{DM,FW}.
Note, in this case, each fractional brane carries
only a half of the zero-brane charge $v (\CB_p) = (0,0,1)$,
\be
v(\CB_+) + v(\CB_-) = v (\CB_p) \,.
\label{bbbpm}
\ee
More generally, in the equivariant category $\CD^b_{\Gamma} (\C^2)$,
the zero-brane $\CB_p$ corresponds to $\varrho_{{\rm reg}} \otimes \CO_p$,
where $\varrho_{{\rm reg}}$ is the regular representation of $\Gamma$.
The representation $\varrho_{{\rm reg}}$ is reducible and,
according to a fundamental theorem of finite group theory,
decomposes as $\varrho_{{\rm reg}} = \oplus_{i} d_i \varrho_i$,
where $d_i = \dim (\varrho_i)$ and $\dim (\varrho_{{\rm reg}}) = |\Gamma|$.
Therefore, in terms of the fractional branes \eqref{sbranes}, we have
\be
\CB_p \; = \; \bigoplus_{\varrho_i \in {{\rm Irr}} (\Gamma)} d_i \CB_i \,.
\label{bpviabi}
\ee
This provides us with a good supply of $(B,B,B)$ branes localized at the orbifold
singularities of $\tilde Y$.

In order to describe $(B,B,B)$ branes on $\tilde Y$ for generic values of $\alpha$,
one needs to understand what happens to the fractional branes $\CB_i$
under the minimal resolution of the Kleinian quotient singularity $\C^2/\Gamma$.
The answer comes from the following equivalence (the derived McKay correspondence)
\be
\CD^b (X) \; \cong \; \CD^b_{\Gamma} (\C^2) \,,
\label{dmckay}
\ee
where $X$ denotes the minimal resolution.
According to \cite{KV}, in the derived category of $X$,
the simple objects \eqref{sbranes} are represented by
\begin{eqnarray}
\phantom{\Big(x\Big)} \CB_1 & = & \CO_{\sum d_i D_i} \,, \label{sbrresol} \\
\phantom{\Big(x\Big)} \CB_i & = & \CO_{D_i} (-1)[1] \,, \qquad i \ne 1 \,, \nonumber
\end{eqnarray}
where $D_i$ are the exceptional divisors.
In particular, in the derived category of $X$ it is easy to see that
each fractional brane $\CB_i$ is a spherical object, {\it i.e.}
\be
\Ext^*_{X} (\CB_i, \CB_i) \; \cong \; H^* (\cp^1, \C) \,.
\ee
This gives yet another reason to identify the fractional branes on $\tilde Y$
with the duals of Lagrangian $A$-branes supported on irreducible
components of the singular fiber \eqref{nilconetoy},
since each component is a copy of $\cp^1$, {\it cf.} \cite{FW}.
(Remember, the first hint came from \eqref{bccrankpunct},
which was then further supported by \eqref{bbbpm}
and the fact that $\tilde Y$ has orbifold singularities.)

In order to identify the mirror $(B,B,B)$ branes more carefully,
it is convenient to start at $\alpha = \frac{1}{2}$.
As we explained earlier, at this special value of $\alpha$ the hypersurface \eqref{mirrorcubic} develops
a singularity of type $D_4$ which, luckily, is also a quotient singularity $\C^2 / \Gamma$
by the binary dihedral group $\Gamma = BD_8$,
whose action on $\C^2$ is generated by the two elements,
\be
\gamma_1 \; = \;
\begin{pmatrix}
\xi & 0 \\
0 & \xi^{-1}
\end{pmatrix}
\qquad {\rm and} \qquad
\gamma_2 \; = \;
\begin{pmatrix}
0 & 1 \\
-1 & 0
\end{pmatrix} \,,
\ee
with $\xi = \exp (\pi i / 2)$. The group $\Gamma = BD_8$
has one $2$-dimensional irreducible representation $\varrho_0$
and four $1$-dimensional irreducible representations $\varrho_i$, $i = 1, \ldots, 4$
(which altogether can be put into three $2$-dimensional
representations, two of which are reducible).
Therefore, from \eqref{bpviabi} we find that, at the $D_4$ singularity on $\tilde Y$,
the zero-brane $\CB_p = \Phi_{{\rm mirror}} (\CB_{\FF})$ is reducible and decomposes as
\be
\CB_p \; = \; 2 \CB_0 \oplus \bigoplus_{i=1}^4 \CB_i \,.
\ee
Comparing this to \eqref{fmhomoltoy}, one is led to identify
$\CB_0$ with $\tCB'$ and $\CB_i$, $i=1, \ldots, 4$ with the branes
mirror to the four $(B,A,A)$ branes supported on the exceptional divisors $D_i \subset Y$.

Indeed, since all of these $(B,A,A)$ branes are supported on irreducible components
of the singular fiber \eqref{nilconetoy} over $b=0$, we expect the mirror
$(B,B,B)$ branes 
to be also supported
on various components of the singular fiber of $\tilde Y$.
In particular, their Chern characters must be linear combinations of
the Poincar\'e duals of the $2$-cycles\footnote{The homology classes
of $\tilde M$ and $\tilde D_i$ generate $H_2 (\tilde Y) \cong \Z^5$
with the intersection form \eqref{torusintersection} and obey an analog
of the relation \eqref{fmhomoltoy}.}
$\tilde M, \tilde D_1, \ldots, \tilde D_4$
and, of course, the class of a point $\ch (\CB_p) = - p$.
Specifically, from \eqref{sbrresol} we find
\be
\ch (\tCB') \; = \;
\frac{1}{2} \left( - \tilde \FF + \tilde D_1 + \tilde D_2 + \tilde D_3 + \tilde D_4 \right)
\label{chbmddual}
\ee
%
According to \eqref{etet} and \eqref{bmukaicharge},
as a function of $\alpha$ the $0$-brane charge of the brane $\tCB' = \CB_0$
is equal to $- (\eta_1 + \eta_2 + \eta_3 + \eta_4) = \frac{1}{2} - \alpha$.
This fact plays an important role in the application to the Verlinde formula.
Indeed, together with \eqref{bccrankpunct},
it determines the leading contribution\footnote{The subleading
constant term was already discussed in \eqref{dimhfirstguess}.}
to the dimension of $\CH = \Ext^*_{\tilde Y} (\tBcc,\tCB')$
which, according to \eqref{GRRtheorem}, is given by
\be
\dim \CH \; = \;
\int_{\tilde Y} \ch (\tBcc)^* \wedge \ch (\tCB') \wedge {\rm Td} (\tilde Y)
\; = \; k - \lambda + 1
\ee
and matches exactly \eqref{verlindetorus} if for $\tBcc$
we take the sheaf on $\tilde Y$ that {\it descends} from
the hyperholomorphic sheaf on $\C^* \times \C^*$ described in section \ref{sec:torus}.

Indeed, in section \ref{sec:torus} we discussed a hyperholomorphic sheaf on $\C^* \times \C^*$
with the Chern character \eqref{chbccmirrortoy},
which is invariant under the $\Z_2$ action.
Hence, it can be thought of as
a $\Z_2$-equivariant coherent sheaf with the trivial $\Z_2$-equivariant structure
that defines a coherent sheaf $\tBcc$ on $\tilde Y$ via the functor
\be
\Phi ~:~~~ \CD^b_{\Z_2} (\C^* \times \C^*) \;\overset{\sim}{\longrightarrow}\; \CD^b (\tilde Y) \,,
\label{bkrmap}
\ee
which is an equivalence \cite{BKR} between the bounded derived category of coherent sheaves on $\tilde Y$
and the bounded derived category of $\Z_2$-equivariant coherent sheaves on $\C^* \times \C^*$.
To be more specific,
the functor \eqref{bkrmap} is obtained by considering the following commutative diagram 
\be
\begin{array}{ccc}
~\CZ & \;\overset{p}{\longrightarrow}\; & \C^* \times \C^* \\
{}^{\kappa} \big\downarrow & \;\phantom{\bigg\downarrow}\; & \big\downarrow {}^{\kappa'} \\
~\tilde Y & \;\overset{p'}{\longrightarrow}\; & ( \C^* \times \C^* ) / \Z_2
\end{array}
\label{bkrdiagr}
\ee
in which $p$ and $p'$ are birational,
$\kappa$ and $\kappa'$ are finite of degree $2$,
and $\kappa$ is flat
(see \cite{BKR} for further details and examples).

Finally, we note that, with little extra work, one can extend the analysis
in the present section to reproduce the Verlinde formula for the four-punctured sphere.
In that case, the mirror manifolds $Y$ and $\tilde Y$ are also cubic surfaces,
similar to \eqref{toruscubic} and \eqref{mirrorcubic},
which in general depend on four holonomy parameters
$\alpha_1$, $\alpha_2$, $\alpha_3$, and $\alpha_4$
(that we already found useful in our discussion here).
They exhibit a more elaborate structure of singularities
({\it cf.} Figure \ref{fig:singpts}) and a rich spectrum
of branes that account for the intricate structure
of the Verlinde formula \eqref{verlindeparabolic}.


\bigskip
{\it Acknowledgments~}
I would like to thank the organizing committee of the Takagi Lectures
for inviting me,
and to acknowledge helpful discussions with E.~Frenkel, T.~Hausel, and E.~Witten.
I also would like to thank E.~Witten for collaboration on
the $A$-model approach to quantization reviewed in section \ref{sec:qart}.
This work is supported in part by DOE Grant DE-FG03-92-ER40701
and in part by NSF Grant PHY-0757647.
Opinions and conclusions expressed here are those of the author
and do not necessarily reflect the views of funding agencies.


\bibliographystyle{amsalpha}

\end{document}